\documentclass[submission,Phys]{SciPost}
\usepackage{amssymb}
\usepackage{float}
\usepackage{subfig}
\usepackage{color}
\usepackage[ngerman, english]{babel}
\usepackage[utf8]{inputenc}
\usepackage{comment}
\usepackage{amsthm}
\usepackage{mathrsfs}
\usepackage{amsbsy}
\usepackage{amsfonts}
\usepackage{physics}
\usepackage{cleveref}
\usepackage{comment}
\usepackage{booktabs}
\usepackage{cancel}
\usepackage{xurl}
\hypersetup{breaklinks=true}
\usepackage[normalem]{ulem}

\definecolor{darkgreen}{rgb}{0,0.5,0}

\newcommand{\torque}{\mathcal{T}}

\newcommand{\etprpt}{\tilde{\eta}_\perp}
\newcommand{\etpart}{\tilde{\eta}_\parallel}
\newcommand{\freq}{\varpi}

\newcommand{\ave}[1]{\left\langle #1 \right\rangle}
 \newcommand{\eqn}[1]{Eq.\,(\ref{#1})}

\newcommand{\vn}[1]{{ \cos\left( #1 \left(\phi-\phi_{0n}\right) \right)  }}

\allowdisplaybreaks

\begin{document}

\begin{center}
{\Large 
\textbf{Hall Viscosity in the Quark-Gluon Plasma}
}
\end{center}

\begin{center}
S. Mondkar\textsuperscript{1,2}, 
G. Torrieri\textsuperscript{3},
M. Kaminski\textsuperscript{4},
R. Meyer\textsuperscript{5,6},$^\ast$
\end{center}

\begin{center}
{\bf 1} Harish-Chandra Research Institute, Chhatnag Road, Jhunsi, Prayagraj 211 019, India \\
{\bf 2} Homi Bhabha National Institute, Training School Complex, Anushakti Nagar, Mumbai 400 094, India \\
{\bf 3}  Instituto de Fisica Gleb Wataghin, University of Campinas, Campinas, Brazil\\
{\bf 4} Department of Physics and Astronomy, University of Alabama, 514 University Boulevard, Tuscaloosa, AL 35487, USA
\\
{\bf 5} Institute for Theoretical Physics and Astrophysics, Julius-Maximilians Universität Würzburg, and Würzburg-Dresden Cluster of Excellence ctd.qmat, 97074 Würzburg, Germany
\\
{\bf 6} Shanghai Institute for Mathematics and Interdisciplinary Sciences (SIMIS), Shanghai, 200433, China
\\
* rene.meyer@physik.uni-wuerzburg.de
\end{center}

\begin{center}
\today
\end{center}

\abstract{We study the Hall viscosity of the quark–gluon plasma (QGP) created in non-central heavy-ion collisions. In the presence of a strong magnetic field or vorticity, rotational symmetry is broken from $O(3)$ to $O(2)$, allowing for two independent Hall viscosities associated with shear deformations transverse and parallel to the symmetry-breaking direction. We find the corresponding constitutive relations by extending the kinetic-theory mechanism to three spatial dimensions and provide parametric estimates of the Hall viscosities under realistic QGP conditions. Both kinetic-theory and holographic estimates indicate that Hall viscosities are comparable in magnitude to the shear viscosity at zero magnetic field. We further show that Hall viscous stresses at hydrodynamic initialization can be as large as standard viscous corrections and identify observable consequences in flow and event-plane correlations.
}

\vspace{10pt}
\noindent\rule{\textwidth}{1pt}
\tableofcontents\thispagestyle{fancy}
\noindent\rule{\textwidth}{1pt}
\vspace{10pt}

\section{Introduction}

\paragraph{} The quark-gluon plasma (QGP) created in ultra-relativistic heavy-ion collision (HIC) experiments behaves as a nearly perfect fluid with a small specific shear viscosity. Relativistic hydrodynamics has emerged as a remarkably successful framework for describing the spacetime evolution of the QGP at intermediate times. A broad set of collective observables, including anisotropic flow coefficients $v_n$, their event-by-event fluctuations and multiparticle cumulants, as well as long-range rapidity correlations, are quantitatively described by viscous relativistic hydrodynamics, initialled by Glauber or color glass initial conditions and converted into particle distributions using an isentropic Cooper-Frye ansatz~\cite{Heinz:2013th, Gale:2013da, Bernhard:2015hxa}.
However, the exact parameters characterizing this fluid are  subject to both theoretical and phenomenological studies~\cite{theorev,phenorev}, the latter employing state-of-the-art machine learning techniques~\cite{Zhou:2023pti}.

This effort is justified by the fact that not all possible transport coefficients have yet been studied. There is growing evidence that strong and rapidly varying electromagnetic fields are present in non-central HICs reaching peak values of $10^{18}- 10^{20}$ G at RHIC/LHC energies~\cite{Shen:2025unr, PhysRevC.107.034901, Skokov:2009qp,directed, Jiang:2024bez}. These are accompanied by sizable vorticity \cite{STAR:2017ckg} associated with the initial angular momentum of the system. However, the full constitutive relations of hydrodynamic fields in the presence of magnetic fields and vorticity are still not fully available \cite{magnrev,theorev}. The situation is both more complicated and interesting, as magnetic fields and vorticity break fundamental symmetries such as isotropy, allowing for a much greater variety of transport coefficients. Furthermore, we know that topological features of QCD \cite{chiraleffect} induce quantum anomalies and hence further transport coefficients \cite{Son:2009tf}, in the presence of both magnetic fields and vorticity. The effective theory of hydrodynamics, including all these effects, is likely to be much richer than that considered in \cite{magnrev,theorev}, and the sort of broad data analysed in \cite{phenorev} could well contain the effects of such hitherto undiscovered terms.

\begin{figure}[t]
  \centering
  \subfloat[\label{figvort}]{
    \includegraphics[width=0.35\linewidth]{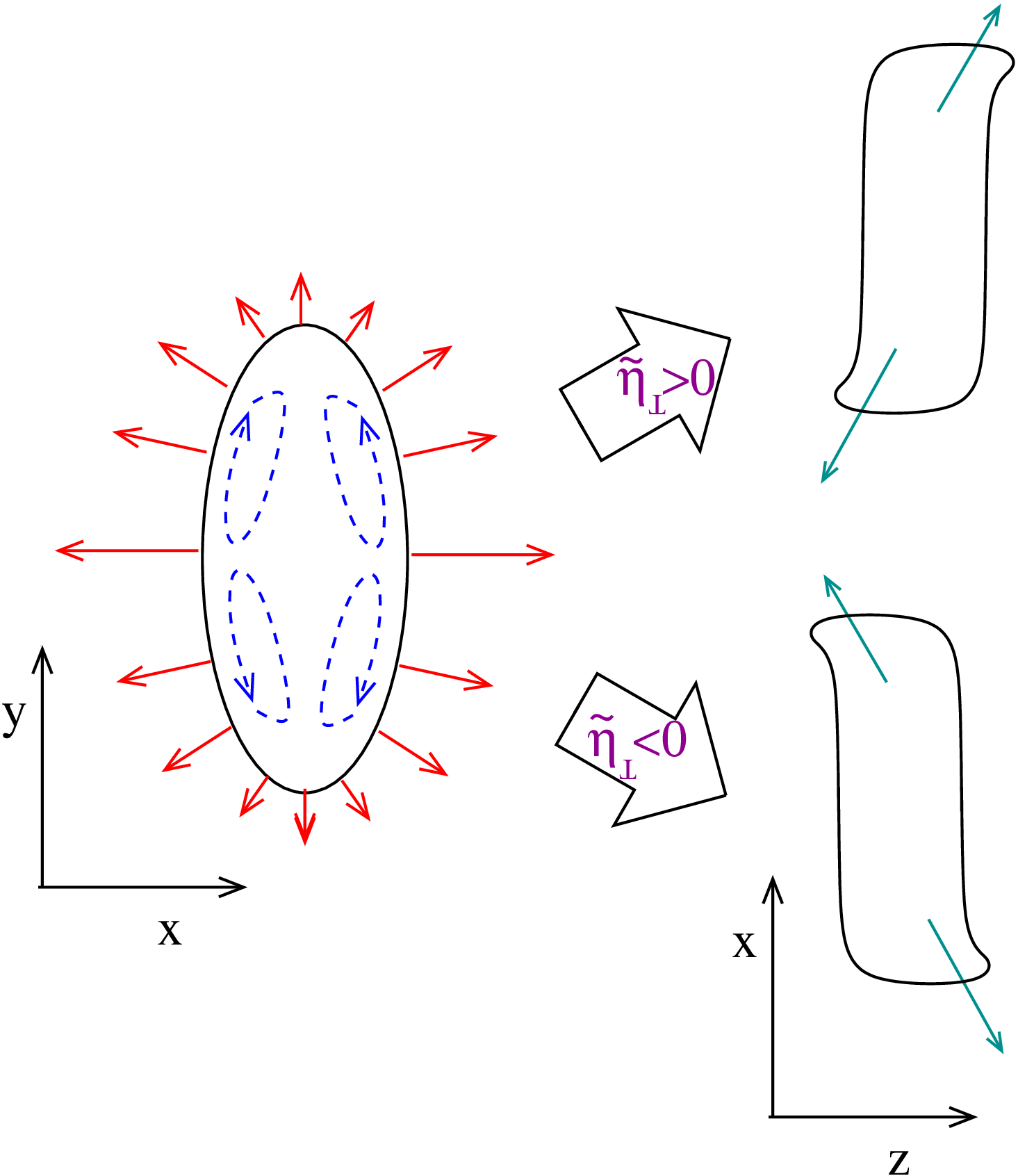}
  }\hfill
  \subfloat[\label{figvortlong}]{
    \includegraphics[width=0.45\linewidth]{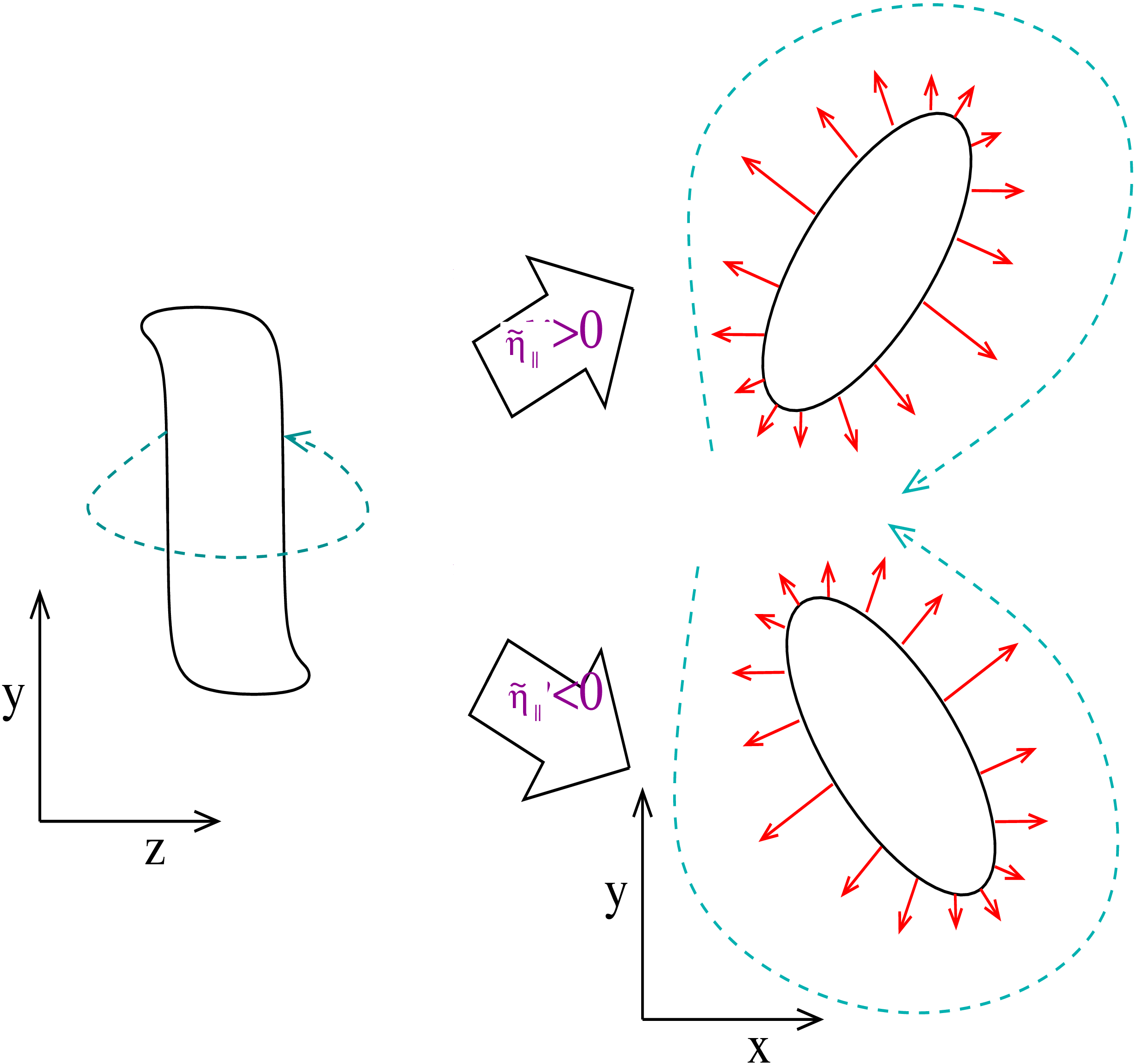}
  }
  \caption{(a) A qualitative illustration of the 
effect of the transverse Hall viscosity $\tilde\eta_\perp$ on a QGP fireball with elliptic flow. This viscosity couples vorticity to azimuthal shear, $T_{xx}-T_{zz}$, hence causing an elongation of the fireball due to longitudinal polarization. (b) A qualitative illustration of the 
effect the longitudinal Hall viscosity $\tilde\eta_\parallel$ on a QGP fireball with transverse polarization. This viscosity couples azimuthal to longitudinal shear, causing the fireball to spin, i.e. rotate perpendicular to the longitudinal vorticity.}
  \label{fig:vort-n-vortlong}
\end{figure}

\begin{figure}[t]
\centering
\includegraphics[width=0.75\textwidth]{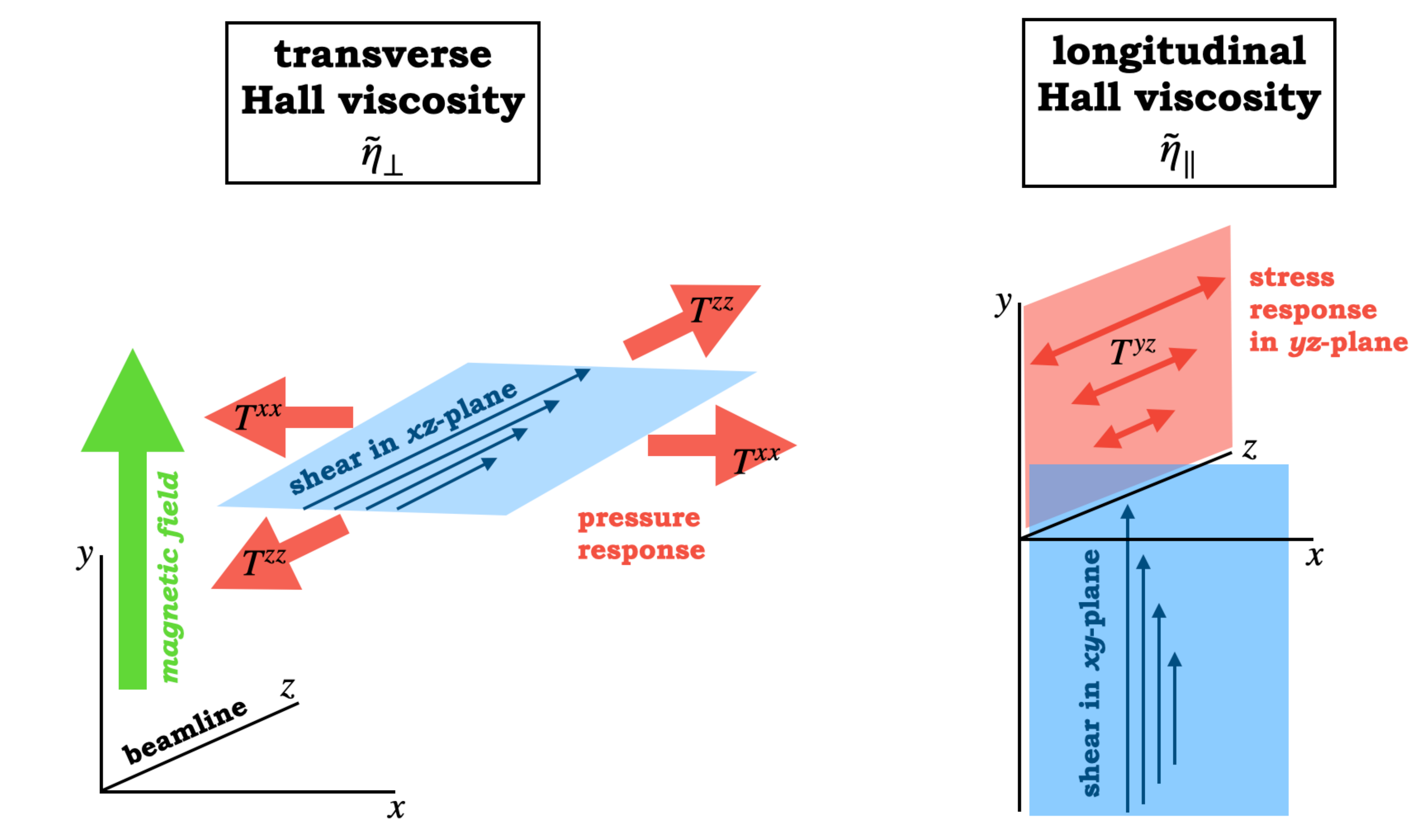}
\caption{\label{fig:HallViscosities}  
Illustration of the distinct effect of the two Hall viscosities on the stress-energy tensor components $T^{\mu\nu}$ in response to different shear flows. The magnetic field points along the $y$-direction, the beamline is along $z$, and the fluid flow has been chosen to have a shear in either the $xz$-plane or the $xy$-plane.}
\end{figure}

In this work, we concentrate on a particular transport coefficient, the Hall viscosity, which, remarkably, results from such a rotational symmetry breaking and, as we show, leads to a unique experimental signature, c.f.~Fig.\ref{fig:vort-n-vortlong} for an illustration. Hall viscosity \cite{Avron_1995,avron1998oddviscosity,Hoyos:2014pba} is a non-dissipative transport coefficient leading to novel hydrodynamic effects. From kinetic theory
in two-dimensional Dirac semimetals like graphene, it is known that a magnetic field generates a Hall viscosity, $\etprpt$, in a plane perpendicular to it~\cite{alekseev2016negative}.  
In three spatial dimensions, however, the structure of Hall viscous transport is richer and more subtle. Once rotational symmetry is broken from $O(3)$ to $O(2)$ by a background field such as the direction of a magnetic field or vorticity, two  independent Hall viscosity coefficients are allowed, corresponding to shear deformations parallel and perpendicular to the symmetry-breaking direction~\cite{LL,Landsteiner:2016stv,Ammon:2020rvg}. This is illustrated in Fig.~\ref{fig:HallViscosities} for a magnetic field breaking the rotational symmetry. 
These two Hall viscosities are related through Kubo formulae~\cite{Kubo:1957mj} to the retarded correlator of the off-diagonal components of the stress-energy tensor $T^{\mu\nu}$ as ~\cite{Landsteiner:2016stv,Landsteiner:2016led,Hernandez:2017mch,Ammon:2020rvg} 
\begin{equation} \label{eq:KuboFormulaeHallViscosities}
    \tilde\eta_{\parallel}=\lim_{\omega \rightarrow 0} \frac{\Im \langle T^{yz} T^{xy} \rangle(\omega)}{\omega}\, ,
    \quad
    \tilde\eta_{\perp}=\lim_{\omega \rightarrow 0} \frac{\Im \langle T^{xz} (T^{xx}-T^{zz}) \rangle(\omega)}{\omega}\, ,
\end{equation}
evaluated at vanishing spatial momentum and in the limit of vanishing frequency $\omega\to 0$.  
Despite being allowed by symmetry, such Hall viscous terms have received comparatively little attention in the context of QGP phenomenology.

The primary goal of this work is to systematically analyze Hall viscosity in the QGP and to assess its potential phenomenological relevance in HICs. We first discuss how in non-relativistic hydrodynamics with broken rotational symmetry, the stress tensor admits two distinct Hall viscosity coefficients, which we denote as $\etprpt$ and $\etpart$. This is accomplished by extending a simple kinetic-theory mechanism originally proposed by Alekseev for two-dimensional charged fluids to three dimensions, demonstrating how Hall viscosity arises from Lorentz-force-induced distortions of the distribution function. We derive the corresponding constitutive relations from the kinetic theory mechanism of \cite{alekseev2016negative}, and estimate the size of $\tilde\eta_{\parallel}$ and $\tilde\eta_{\perp}$.  Complementing this weak-coupling analysis, we also discuss estimates of Hall viscosity at strong coupling using holographic methods.

Finally, we focus on the implications of Hall viscosity for the early-time dynamics of the QGP. Rather than attempting a full magnetohydrodynamic simulation, we provide controlled order-of-magnitude estimates of Hall viscous stress corrections at hydrodynamic initialization time, using realistic velocity gradients appropriate for HICs. We outline how Hall viscosity may influence the evolution of collective flow in HICs and identify an experimental observable sensitive to the Hall viscous effects.

Before we proceed, a comment on the convention of the coordinate axes is in order. In the heavy-ion literature, the most commonly adopted convention is as follows: The beam axis direction is along the z-axis, and the impact parameter direction is along the $x$-axis. Therefore, the $x,z$-plane constitutes the reaction plane. The $y$-axis direction is the out-of-plane direction. For off-central HICs, the overlap region of the two colliding nuclei at the point of closest approach is an almond-shaped region with the short axis of the almond along the impact parameter direction ($x$-axis) and the long axis of the almond along the out-of-plane direction ($y$-axis). This initial spatial anisotropy in the $x$ and $y$ directions gives rise to pressure anisotropy in these two directions, resulting in the rapid expansion of the fireball in the $x$-direction compared to its expansion in the $y$-direction. This pressure anisotropy gives rise to the flow coefficient $v_2$. The magnetic field generated in the collision of the two nuclei is initially oriented perpendicular to the reaction plane, that is, in the $y$-direction~\cite{Skokov:2009qp, PhysRevC.85.044907, Bzdak:2011yy, PhysRevC.83.054911}. Additionally, the axis of rotation of the fireball is also oriented perpendicular to the reaction plane, so also in the $y$-direction~\cite{Jiang:2016woz, Becattini:2020ngo, Rath:2024vyq}.

Before proceeding, we emphasize the scope of the present work. Our analysis is not a full relativistic magneto-hydrodynamic simulation of the QGP, nor does it include the dynamical evolution of electromagnetic fields. Instead, our goal is to establish the existence, parametric magnitude, and phenomenological relevance of Hall viscous transport in the QGP through controlled analytic estimates. The results presented here should therefore be viewed as order-of-magnitude benchmarks that motivate future quantitative simulations incorporating Hall viscosity.
In this context, the main difference between relativistic and non-relativistic hydrodynamics is that in the latter, compressibility and dissipative transport coefficients are strongly correlated while in the former compressibility has a lower bound due to causality \cite{liaokoch}.
However, the model that we use \cite{alekseev2016negative} is in the compressible regime.  Because of this, we are confident that relativistic and non-relativistic results are well within an order of magnitude, and hence using non-relativistic ansatze is sufficient for our purposes. Furthermore, our estimates for the Hall viscosities from holographic states, which are fully relativistic and compressible, yield an estimate of the same order of magnitude as the one derived from our non-relativstic Ansatz.

The main new results of this work are threefold: (i) a systematic derivation of the  
non-relativistic 
Hall viscous terms in the constitutive relations in three spatial dimensions with broken rotational symmetry from kinetic theory, (ii) the first quantitative estimates of Hall viscosities for QGP-relevant parameters using both kinetic theory and holography, and (iii) the identification of experimentally accessible signatures of Hall viscosity in heavy-ion collisions.

The paper is organized as follows. In Sec.~\ref {sec:HallViscosityQGP::subsec::Alekseev3D}, we derive the constitutive relations for three-dimensional non-relativistic hydrodynamics in the presence of magnetic fields and derive the expressions for Hall viscosities. In Sec.~\ref{sec:HallViscosityQGP::subsec:OOM-estimate-KT} and~\ref{subsec:OOM-estimate-holo}, we present kinetic-theory and holographic estimates of the Hall viscosity coefficients, respectively. In Sec.~\ref{sec:HallObservables}, we evaluate the size of Hall viscous corrections under QGP conditions and discuss their phenomenological implications. We conclude in Sec.~\ref{sec:ConcOutlook} with a summary and outlook.

\section{Generating Hall viscosity in the QGP}\label{sec:HallViscosityQGP}
In this section, we will explain our mechanism of how Hall viscosity is generated in the magnetic fields present in HICs, and estimate the order of magnitude of the resulting Hall viscosities both in the perturbative regime as well as from holography. 

\subsection{Hall viscosities in three-dimensional non-relativistic hydrodynamics}\label{sec:HallViscosityQGP::subsec::Alekseev3D}
\paragraph{}
In three spatial dimensions, Hall viscosity is a genuine tensor component of the viscosity tensor \cite{Hoyos:2014pba}, and hence vanishes identically in the presence of rotational invariance even if  time-reversal symmetry is broken~\cite{avron1998oddviscosity}. However, the magnetic field breaks both  time-reversal  and  rotational invariance, and therefore, non-zero Hall viscosity can be obtained in the presence of a magnetic field. In this subsection, we extend the  mechanism proposed by Alekseev~\cite{alekseev2016negative} for the generation of two-dimensional Hall viscosity, which we review in App.~\ref{App:A}, to non-relativistic kinetic theory in three dimensions. We will show that in three dimensions two Hall viscosities emerge in the presence of an external magnetic field, one parallel and one in the plane perpendicular to the magnetic field. 

{Alekseev's mechanism \cite{alekseev2016negative} relies on kinetic theory, which itself relies on the existence of quasiparticles in the described medium. In the QGP, the relevant quasiparticles to which the magnetic field couples are in-medium quarks, and in the following we denote by 
$\tau_2$ the relaxation time associated with the second moments of the quasiparticle distribution function.\footnote{In the solid state systems considered in \cite{alekseev2016negative}, the relaxation time of the first moment of the distribution function will be relevant to include the effect of momentum relaxation. In the QGP, momentum is conserved even in medium, and we hence do not include such terms in our kinetic theory treatment.} The reason that we can apply  non-relativistic kinetic theory to the QGP is that the magnetic field induced Hall viscosity stems from the coupling of the magnetic field to the quarks. Since advective forces on a cell of size of the mean free path $\lambda$ are of order $\frac{\lambda}{e} \frac{dp}{dx}$, these forces will not induce large comoving velocities.  Hence, in a comoving cell frame, the quarks, with their thermally dressed masses being ${\cal O} (600 MeV)$, can be treated non-relativistically for a qualitative estimate of the Hall viscosities and the effects induced by them.} 
The viscous stress tensor is given by $\Pi_{ij} := m \langle v_i v_j \rangle$, where $\textbf{v} = (v_x, v_y, v_z)$ is the three-dimensional velocity of a single quasiparticle and the angular brackets stand for averaging over the quasiparticle velocity distribution at a given spatial location $\textbf{r} = (x,y,z)$. The equation of motion for the hydrodynamic velocity $V_i = \langle v_i \rangle$ in the absence of magnetic field is

\begin{equation} \label{eq:hydroeom}
    m \frac{\partial V_i}{\partial t} = - \frac{\partial \Pi_{ij}}{\partial x_j} + e E_i \, ,
\end{equation}
where $E_i$ is the external electric field, and we have assumed Einstein summation convention. In M\"uller-Israel-Stewart theory,  $\tau_2$ corresponds to the M\"uller-Israel-Stewart  relaxation time~\cite{Muller:1967zza,Israel:1976tn,Israel:1979wp}. At a time scale much greater than $\tau_2$, the expression for $\Pi_{ij}$ is given by its steady-state value
\begin{equation}\label{eq:Pi0}
    \Pi_{ij} = \Pi^{(0)}_{ij} = - \frac{m}{\varrho} \eta_0 V_{ij}, \qquad V_{ij} = \frac{\partial V_i}{ \partial x_j} + \frac{\partial V_j}{\partial x_i} \, ,
\end{equation}
where $m$ is the quasiparticle mass and $\varrho$ is the quasiparticle mass density. $\eta_0$ is the dynamical viscosity with the same dimensions as the entropy density, $s$, in the natural units. In the non-relativistic literature, the kinematic viscosity, $\nu$, is often used, but there is no meaningful relativistic generalization of it.  Thus, in this work, we shall use the usual dynamical viscosity. The two are related by the expression
\begin{align}\label{eq:kinvisc}
    \nu = \frac{\eta_0}{s T + \varrho} \, ,
\end{align}
where $T$ denotes the temperature. In the non-relativistic limit $\varrho \gg s T$, which implies $\nu \approx  \eta_0/ \varrho$. In the ultrarelativistic limit, \eqref{eq:kinvisc} is still valid, with $s T + \varrho \approx sT$.

$\Pi_{ij}$ relaxes to its steady-state value $\Pi^{(0)}_{ij}$ during the time $\tau_2$ according to the Drude-like equation
\begin{equation}\label{eq:Pieom}
    \frac{\partial \Pi_{ij}}{\partial t} = - \frac{1}{\tau_2} \left( \Pi_{ij} - \Pi^{(0)}_{ij} \right) \, .
\end{equation}

\noindent Introducing an external constant magnetic field, $B_i$, shifts the steady-state value of $\Pi_{ij}$ from $\Pi^{(0)}_{ij}$. 
Without loss of generality, we choose the magnetic field to point along the $\hat{\textbf{y}}$ direction. This choice explicitly breaks the spatial $O(3)$ rotational
symmetry down to an $O(2)$ symmetry in the plane transverse to $\mathbf B = B  \hat{\textbf{y}}$.
In the presence of a magnetic field, the time evolution of $V_i$ and $\Pi_{ij}$
is influenced not only by collisions and the electric field, but also by the magnetic component of the Lorentz force.  Consequently, the evolution equations for  $\partial V_i / \partial t$ and $\partial \Pi_{ij}/\partial t$ acquire additional magnetic-field–induced contributions.
Explicitly, the magnetic terms entering the equations of motion~\eqref{eq:hydroeom} and~\eqref{eq:Pieom} take the form~\cite{alekseev2016negative}
\begin{align}
    \left( \frac{\partial V_i}{\partial t} \right)_{B} = & \omega_c \epsilon_{yik} V_k \label{eq:mag-hydroeom} \, ,\\
    \left( \frac{\partial \Pi_{ij} }{\partial t}  \right)_{B} = &   \omega_c \left( \epsilon_{yik} \Pi_{kj} + \epsilon_{yjk} \Pi_{ik} \right) \label{eq:mag-Pieom}\, ,
\end{align}
where $\omega_c = q B/m$ is the cyclotron frequency and $q$ denotes the electric charge of quasiparticles.
We have assumed that the electromagnetic fields, appearing implicitly in $\omega_c$, do not alter the particle mass significantly. If they would do so, the non-relativistic approximation would be not applicable in any case, and these effects would also show up in the resulting transport coefficients.
The terms \eqref{eq:mag-hydroeom} and \eqref{eq:mag-Pieom} are added to the right-hand side of equations  \eqref{eq:hydroeom} and \eqref{eq:Pieom}, respectively. 

We work in the hydrodynamic Navier--Stokes regime
and focus on the steady state, $\partial_t \Pi_{ij}=0$ appropriate for timescales long compared to the microscopic relaxation time $\tau_2$. Under these
assumptions, the evolution equation~\eqref{eq:Pieom} reduces to an algebraic relation between $\Pi_{ij}$ and the symmetrized velocity gradients. 
Therefore, in the steady state,  $\partial_t \Pi_{ij}=0$, from the modified equation \eqref{eq:Pieom} in the presence of a magnetic field, we get the following relation
\begin{equation}\label{eq:steady-state}
    \Pi_{ij} - \omega_c \tau_2 \left( \epsilon_{yik} \Pi_{kj} + \epsilon_{yjk} \Pi_{ik} \right) = \Pi^{(0)}_{ij} \, .
\end{equation}
Note that $\Pi_{ij}$ denotes the dissipative part of the stress-energy tensor, obtained after
subtracting the isotropic equilibrium pressure contribution. For incompressible
flow, $\text{div} \textbf{V}:= \partial V_i/ \partial x_i = 0$, and Eq.~\eqref{eq:steady-state} implies that the viscous correction is traceless, $\Pi_{ii} = 0$.
Subsequently, only two of the three diagonal components of $\Pi_{ij}$ are independent. Since $\Pi_{ij}$ is a symmetric rank-2 tensor, there are three independent off-diagonal components. In total, $\Pi_{ij}$ has five independent components: $\Pi_{xx}$, $\Pi_{zz}$, $\Pi_{xz}$, $\Pi_{xy}$, $\Pi_{zy}$. We can get five coupled algebraic equations for these five components from \eqref{eq:steady-state}.
Solving the resulting coupled algebraic equations for the components
of $\Pi_{ij}$, we obtain the following constitutive relations between the stress
tensor and the symmetrized velocity gradients,
$V_{ij}=\partial_i V_j+\partial_j V_i$,

\begin{subequations}\label{eq:const-3d}
\begin{align}
 -\frac{\Pi_{xx}}{(m/\varrho)} = &  \frac{\eta_0}{2}  \left(V_{xx} + V_{zz} \right) + \frac{\eta_\perp}{2} \left( V_{xx} - V_{zz} \right) -  \tilde{\eta}_\perp V_{xz}   \label{eq:const-3d:sub1} \, ,\\
    -\frac{\Pi_{zz}}{(m/\varrho)} = & \frac{\eta_0}{2}  \left(V_{xx} + V_{zz} \right) - \frac{\eta_\perp}{2} \left( V_{xx} - V_{zz} \right) +  \tilde{\eta}_\perp V_{xz} \ \label{eq:const-3d:sub2} \, ,\\
    -\frac{\Pi_{xz}}{(m/\varrho)} = &  \frac{\tilde{\eta}_\perp}{2}  \left( V_{xx} - V_{zz}\right) +  \eta_\perp  V_{xz} \label{eq:const-3d:sub3} \, , \\
    -\frac{\Pi_{xy}}{(m/\varrho)} = &  \eta_\parallel V_{xy} -  \frac{\tilde{\eta}_\parallel}{2} V_{zy} \label{eq:const-3d:sub4} \, , \\
    -\frac{\Pi_{zy}}{(m/\varrho)}= &  \eta_\parallel V_{zy} +  \frac{\tilde{\eta}_\parallel}{2} V_{xy} \label{eq:const-3d:sub5} \, .
\end{align} 
\end{subequations}
The residual $O(2)$ symmetry about the magnetic-field direction allows two
independent shear viscosities and two independent Hall viscosities.
$\eta_{\perp}$ and $\eta_{\parallel}$ denote the anisotropic shear
viscosities. Shear viscosities are time-reversal invariant and are dissipative transport coefficients that contribute to the irreversible entropy production.
In contrast, $\tilde\eta_{\perp}$ and
$\tilde\eta_{\parallel}$ are Hall viscosities, which are odd under time-reversal
and encode non-dissipative responses induced by the magnetic field.
These four viscosities can be expressed in terms of zero magnetic field viscosity $\eta_0$ as

\begin{equation}\label{eq:visc-B}
    \eta_\perp = \frac{\eta_0}{1 + \freq^2}, \quad \eta_\parallel = \frac{4}{4 + \freq^2} \eta_0, \quad  \tilde{\eta}_\perp = \frac{\freq}{1 + \freq^2} \eta_0, \quad \tilde{\eta}_{\parallel} = \frac{ 4 \freq }{4 + \freq^2} \eta_0 \, ,
\end{equation}
where $\varpi := 2 \omega_c \tau_2$. As a consistency check, in the zero-field limit $\freq\to 0$ the Hall viscosities
$\tilde\eta_{\perp}$ and $\tilde\eta_{\parallel}$ vanish, while the shear
viscosities reduce to their isotropic value $\eta_0$ as required by the restoration of
full rotational symmetry.

The existence of two independent Hall viscosities follows purely from symmetry considerations once rotational invariance is broken from $O(3)$ to $O(2)$ and time-reversal symmetry is violated. This symmetry-breaking pattern is model-independent and applies equally to weakly and strongly coupled systems. The quantitative values of these transport coefficients, however, depend on microscopic dynamics and must be estimated using specific theoretical frameworks, which we will do in the next sections.

\subsection{Estimated Hall viscosities from kinetic theory}\label{sec:HallViscosityQGP::subsec:OOM-estimate-KT}

We now turn to estimating the magnitude of the Hall viscosities for the QGP. 
In this subsection we obtain order-of-magnitude estimates for the Hall viscosities of the QGP by extrapolating the kinetic theory expressions derived in Sec.~\ref{sec:HallViscosityQGP::subsec::Alekseev3D} to relativistic plasma conditions. These estimates are not meant to constitute a controlled relativistic kinetic-theory calculation, but rather to provide physical intuition and a benchmark scale for the potential size of Hall viscous effects.

In non-relativistic kinetic theory, the zero magnetic field shear viscosity is given by 

\begin{equation}
    \eta_0 = \frac{1}{3} \sum_i \rho_i \langle p \rangle_i \lambda_i \, ,
\end{equation}

\noindent where $\rho_i$ is the number density of the quasiparticle species $i$, $\langle p \rangle_i$ is the average momentum, $\lambda_i$ is the mean free path. However, for an ultrarelativistic gas, the prefactor changes from $\frac{1}{3}$ to $\frac{4}{15}$~\cite{Mattiello:2010nfi, PhysRevD.31.53}, yielding
\begin{eqnarray}
    \eta_0 = \frac{4}{15} \sum_i \rho_i \langle p \rangle_i \lambda_i \, .
\end{eqnarray}

\noindent The change of factor from $1/3$ to $4/15$ reflects the differences in momentum transfer and particle dynamics in a relativistic regime~\cite{Mattiello:2010nfi, PhysRevD.31.53}. The mean free path is related to the relaxation time $\tau_2$ by 
\begin{equation}
    \lambda_i = v_{i} \tau_2 \approx c \tau_2 \, ,
 \end{equation}
where $v_{i} \approx c $ for ultrarelativistic particles. Thus we get 
\begin{eqnarray}\label{eq:zeroB-visc}
     \eta_0 \approx \frac{4}{15} \sum_i \rho_i \langle p \rangle_i c \tau_2 \, .
\end{eqnarray}
The energy density is given by 
\begin{equation}
     \varepsilon = \sum_i \rho_i \langle E \rangle_i \, .
\end{equation}
For an ultrarelativistic gas, the average energy is approximately equal to the average momentum  $  \langle E \rangle_i \approx \langle p\rangle_i c$. This simplifies the energy density expression to 
\begin{equation} \label{eq:energy-density-rkt}
     \varepsilon  \approx  \sum_i \rho_i \langle p\rangle_i c \, .
\end{equation}
Eq.~\eqref{eq:zeroB-visc} and \eqref{eq:energy-density-rkt} together yield
 \begin{equation}\label{eq:eta-0}
     \eta_0 \approx \frac{4}{15} \varepsilon \tau_2 \, .
 \end{equation}
This relation reflects the fact that, in an ultrarelativistic gas, momentum transport is governed by the energy density and the relaxation time of the second moment of the distribution function, a result well known from relativistic kinetic theory. 
The energy density of the QGP in the high temperature limit is given by 
\begin{equation}
     \varepsilon \approx \frac{\pi^2}{30} g_{\text{eff}} T^4 \, ,
\end{equation}
based on an estimate using the Stefan-Boltzmann law for an ideal gas of quarks and gluons \cite{Blaizot:2001yq, AbouSalem:2015wha, Castorina:2005wi}. $g_{\text{eff}}$ is the effective number of degrees of freedom in QCD, 
\begin{align}
    g_{\text{eff}} = g_g + \frac{7}{8} g_q 
                     = 2 \left( N_c^2 - 1 \right) + \frac{7}{8} 4 N_c N_f 
                     = 37 \, ,
\end{align}
where $g_g = 2 \left( N_c^2 - 1 \right)$ ($N_c$ colors, two polarizations) is the number of gluon degrees of freedom, whereas $g_q = 4 N_c N_f$ ($N_c$ colors, $N_f$ flavors, two spins, quarks and anti-quarks)  is the number of quark degrees of freedom. The contribution from quark degrees of freedom is reduced by a factor of $7/8$ due to fermionic statistics. 
We took $N_c =3$ and $N_f = 2$, i.e., considering only two active light quark flavors for simplicity; including the strange quark would modify numerical prefactors at the $10$–$20 \%$ level without qualitatively affecting the parametric estimates presented below. Thus, we get an approximate energy density of the QGP as  \cite{PhysRevD.31.53}

\begin{equation}\label{eq:e-T}
    \varepsilon \approx 12.17 \; T^4 \, .
\end{equation}
The typical temperature of the QGP is $T \approx 300 \; \text{MeV}$ at RHIC and  LHC \cite{Sas:2022wjp, ALICE:2015xmh, STAR:2024bpc}. 

The holographic prediction of the relaxation time of the QGP is $\tau_2 \approx 0.2-0.3$ fm/c with smooth initial conditions \cite{Balasubramanian:2010ce} (and $0.4-0.6$ fm/c after incorporating initial state fluctuations \cite{Muller:2020ziz}) for the temperatures $ T \approx 300 - 400$ MeV. We take the relaxation time $\tau_2$ to be $\tau_2 \approx 0.4 $ fm/c. Using the relation $1 \; \text{MeV}^{-1} = 197.3 \; \text{fm}/c$, $\tau_2$ can be expressed in terms of the QGP temperature $T$ as $\tau_2 \approx 0.6/  T$ (for $T \approx 300$ MeV). The estimates from holographic models are close to experimentally observed QGP thermalization times. In the present context $\tau_2$ should be viewed as an effective phenomenological relaxation time characterizing the approach to local equilibrium, rather than a quantity derived self-consistently within weakly coupled kinetic theory.

Finally, our estimate for the zero-magnetic field shear viscosity of the QGP~\eqref{eq:eta-0} is 

\begin{equation}\label{eq:eta-0-T}
    \eta_0 \approx \frac{4}{15} 12.17 \,T^4 \frac{0.6}{ T} =  1.95 \; T^3 \, .
\end{equation}
Next, we estimate the entropy density $s$. We have the thermodynamic relation 
\begin{equation}
    s = \frac{\varepsilon + P}{T}  \, .
\end{equation}
Using the conformal equation of state (EoS), $P = \varepsilon/3$, whose validity we discuss below in our regime of parameters, and the  expression for energy density \eqref{eq:e-T}, we get 
\begin{equation}\label{eq:s-T}
    s \approx \frac{4}{3} \frac{1}{T} 12.17 \; T^4 = 16.23 \; T^3 \, .
\end{equation}
Combining \eqref{eq:eta-0-T} and \eqref{eq:s-T}, we get an estimate for the dimensionless ratio $\eta_0/ s \approx 0.12 $, which is of the same order of magnitude as the holographic value for very strongly interacting quantum field theories dual to Einstein Gravity $1/ (4 \pi)$~\cite{Kovtun:2004de,Cremonini:2011iq}.

Let us compare the $\eta_0/s$ we obtained with the $\eta/s $ obtained by Bayesian parameter estimates. In particular, Ref.~\cite{Bernhard:2019bmu} has $\eta/s$ roughly in the range of $0.07 \lesssim  \eta/s \lesssim 0.15$.
This is in the ballpark of our estimated value of $\eta_0/s \approx 0.12$. Note that our estimate of $\eta_0/s$ is independent of temperature whereas the Bayesian estimate of Ref.~\cite{Bernhard:2019bmu} gives temperature dependent $\eta/s$. Recall that in our estimate $\eta_0$ came from using the conformal EoS of QCD with two active quark flavors and using phenomenological estimate for the relaxation time.
So our estimate is only valid at high enough temperatures where QGP EoS can be approximated by the conformal EoS. Ref.~\cite{Bernhard:2019bmu} used a lattice QCD EoS~\cite{PhysRevD.90.094503} for the QGP. As another consistency check, we estimate the quantity $\frac{\tau_2 (\varepsilon + P)}{\eta_0}$ by using the above estimates for $\tau_2 \approx \frac{0.6}{ T}$, $\eta_0 \approx 1.95 \; T^3$, $\varepsilon \approx 12.17 \; T^4$, and the conformal EoS $P = \varepsilon/3$. We get  $\frac{\tau_2 (\varepsilon + P)}{\eta_0}  \approx 5$ which falls within the acceptable range of $5-7$ given in \cite{Czajka:2017bod}, {albeit at the lower end of the range.}

 At $T \approx 300 \; \text{MeV}$, we get $\eta_0 \approx 1.95 \; T^3  = 5.26 \times 10^7 \; \text{MeV}^3$. In order to obtain the estimates for the viscosities of Eq.~\eqref{eq:visc-B} in the presence of external magnetic fields, we first estimate the quantity $\varpi = 2 \omega_c \tau_2$.  $\omega_c$ is the cyclotron frequency and is related, in the non-relativistic limit, to the magnetic field $B$ as $\omega_c = \frac{q B}{m}$. $ B$ is estimated to be $10^{18}$ Gauss at RHIC \cite{PhysRevC.107.034901, Skokov:2009qp,directed} and $10^{19} - 10^{20}$ Gauss at the LHC \cite{Jiang:2024bez}. {For our order-of-magnitude estimate, we take $ B \approx 10^{19}$ Gauss. {We assess the sensitivity to smaller field strengths in
Sec.~\ref{subsec:B-sensitivity}.}
 This estimate applies to early times relevant for the onset of hydrodynamic evolution, where the magnetic field, although already decaying from its initial peak, can still be sizable depending on the electrical conductivity of the medium and can therefore generate Hall viscous stress corrections at initialization.}
 
{Thus, taking the interpretation of experimental results at \cite{directed} as correct, and remembering the scaling of initial magnetic fields with energy then, provided the magnetic field is long-lived enough for Hall viscosity to turn on at LHC energies there is a reasonable chance the effects of Hall viscosity are indeed observable.} 

The effective thermal mass of quarks $m$ that appears in the expression for $\omega_c$ is given by $m \sim g T$ \cite{Seipt:2008wx, Hidaka:2006gd}, where $g$ is related to the dimensionless strong coupling $\alpha_s$ of QCD by $\alpha_s = \frac{g^2}{4 \pi}$. $\alpha_s$ is the QCD running coupling and thus depends on the energy scale. Phenomenological estimates suggest \cite{Shalaby:2012hqh, Mattingly:1993ej, Dai:2020rlu} $\alpha_s (300 \; \text{MeV} ) \approx 0.3$. This gives $g (300 \; \text{MeV}) \approx 1.94$. Using the relations $1$ Gauss $= 1.95 \times 10^{-20} \; \text{GeV}^2$ \cite{Palash-Pal-table}, $B \approx 10^{19}$ Gauss in units of GeV becomes $ B \approx \; 0.195  \; \text{GeV}^2 = 1.95 \times 10^{5} \; \text{MeV}^2$. We can express $B$ in terms of temperature $T$ as $B \approx 2.17 \; T^2$ at $T \approx 300 \; \text{MeV}$.

{Therefore we get $\varpi = 2 \omega_c \tau_2 \approx  2 \frac{q B}{g T} \frac{0.6}{T} \approx 2 \frac{q \, 2.17 \; T^2}{1.94\; T} \frac{0.6}{T}  \approx 1.34 \, q$. Here $q$ denotes the charge of quarks, which is $q_u = 2 e/3$ for up quarks and $q_d = - e/3$ for down quarks, where $e = \sqrt{4 \pi \alpha} \approx 0.303$, $\alpha = 1/137$ being the fine-structure constant. Since we are only interested in order-of-magnitude estimates here, from here on we will choose $q = q_u \approx 0.202$, as choosing $q=q_d$ instead will not change the order-of-magnitude of the estimate. This gives us $\varpi \approx 0.27$.} By plugging this $\varpi$ estimate in Eq.~\eqref{eq:visc-B}, we obtain the following estimates for the viscosities at $T \approx 300$  MeV
{
\begin{align}\label{eq:visc-B-ket}
    \eta_\perp & = \frac{1}{(1 + \varpi^2)} \eta_0  \approx 0.93 \eta_0 \, , \qquad
    \eta_\parallel  = \frac{4}{4 + \varpi^2} \eta_0  \approx 0.98 \eta_0 \, , \nonumber \\
     \tilde{\eta}_\perp & = \frac{\varpi}{ (1 + \varpi^2)} \eta_0  \approx 0.25 \eta_0 \, , \qquad
    \tilde{\eta}_\parallel  = \frac{ 4 \varpi}{4 + \varpi^2} \eta_0  \approx 0.27 \eta_0 \, .
\end{align}
}
Recalling that the zero magnetic field viscosity is $\eta_0 \approx 5.26 \times 10^7 \; \text{MeV}^3$. From \eqref{eq:visc-B-ket}, we observe that the Hall viscosities are roughly the same order of magnitude as the anisotropic shear viscosities.

To assess the sensitivity of our estimates to the effective quasiparticle mass entering the cyclotron frequency, we consider an alternate route. Note that the masses of active u, d, and s quarks in the QGP are very small compared to the QGP temperature scale.
Therefore, in the alternate approach we assume the quarks to be massless (instead of having thermal mass $\sim g T$). The $B$ dependence of the cyclotron frequency for ultra-relativistic massless charged Dirac fermions (electrons and holes) in graphene is given by $\omega_c \propto \sqrt{B} $~\cite{PhysRevLett.98.197403, PhysRevB.79.241309, PhysRevB.95.125402}. We extrapolate this to the massless quarks in the QGP. {For an illustrative alternative estimate, we assume $\omega_c = \sqrt{q_u B}$ instead of $\omega_c = \frac{q_u B}{g T}$, we get $\varpi = 2 \omega_c \tau_2 = 2 \sqrt{q_u B} \tau_2 \approx 2 \sqrt{0.202 \times 2.17 \; T^2} \frac{0.6}{ T}  \approx 0.80$, where we used $B \approx 2.17 \; T^2$, $q_u \approx 0.202$, and $\tau_2 \approx 0.6/T$. With this $\varpi \approx 0.80$, we can provide alternate estimates for various viscosities in Eq.~\eqref{eq:visc-B} as follows}
{
\begin{align}\label{eq:visc-B-ket-massless}
    \eta_\perp & = \frac{1}{(1 + \varpi^2)} \eta_0  \approx 0.61 \eta_0 \, , \qquad
    \eta_\parallel  = \frac{4}{4 + \varpi^2} \eta_0  \approx 0.86 \eta_0 \, , \nonumber \\
     \tilde{\eta}_\perp & = \frac{\varpi}{(1 + \varpi^2)} \eta_0  \approx 0.49 \eta_0 \, , \qquad
    \tilde{\eta}_\parallel  = \frac{ 4 \varpi}{4 + \varpi^2} \eta_0  \approx  0.69 \eta_0 \, .
\end{align}
}
We notice that this estimate \eqref{eq:visc-B-ket-massless} with massless quarks does not differ much from the earlier estimate \eqref{eq:visc-B-ket} with thermal mass quarks.

Taken together, these estimates indicate that both Hall viscosities, $\etprpt$ and $\etpart$, are parametrically comparable to the anisotropic shear viscosities under realistic QGP conditions. This motivates treating Hall viscous contributions on equal footing with dissipative viscous terms when assessing their phenomenological impact.

\subsection{Estimated Hall viscosities from holography}\label{subsec:OOM-estimate-holo}

\paragraph{}
To complement the weak-coupling, kinetic-theory estimates presented in Sec.~\ref{sec:HallViscosityQGP::subsec:OOM-estimate-KT}, we now turn to a strong-coupling estimate of Hall viscosities based on gauge–gravity duality, referring to~\cite{Landsteiner:2016stv,Landsteiner:2016led} and the later~\cite{Ammon:2020rvg}. In particular, we compare to the field-theoretic analysis of chiral hydrodynamics in the presence of strong magnetic fields developed in Ref.~\cite{Ammon:2020rvg}, which  derives the most general relativistic constitutive equations for a relativistic plasma containing chiral fermions in the presence of a strong external magnetic field carrying an anomalous $U(1)$ charge. This field theoretic derivation yields relativistic Kubo formulae for the Hall viscosities. These are the relativistic counterparts to the non-relativistic Hall viscosities introduced in Sec.~\ref{sec:HallViscosityQGP::subsec::Alekseev3D}. Ref.~\cite{Ammon:2020rvg} then provides a controlled computation of transport coefficients for a (3+1)-dimensional relativistic plasma within a holographic model, specifically both Hall viscosities are computed.

The holographic setup consists of a five-dimensional asymptotically AdS charged magnetic black-brane geometry~\cite{DHoker:2009ixq}  within Einstein-Maxwell-Chern-Simons theory, i.e. Einstein gravity with a negative cosmological constant coupled to gauge fields through a Maxwell and a Chern–Simons term, the latter encoding the quantum anomaly of the dual field theory. The boundary theory describes a strongly coupled, charged relativistic fluid at finite temperature and chemical potential, subject to a constant external magnetic field. The magnetic field explicitly breaks spatial rotational symmetry from $O(3)$ down to $O(2)$, thereby allowing for anisotropic transport and parity-odd response coefficients, including Hall viscosities.

As argued above, in (3+1) spacetime dimensions, once rotational symmetry is reduced to $O(2)$, the viscous stress tensor admits two independent non-dissipative, time-reversal-odd viscosity coefficients, corresponding to shear deformations perpendicular and parallel to the symmetry-breaking axis. Within the relativistic setup, these are the two Hall viscosities, $\etprpt$ and $\etpart$, introduced on general hydrodynamic grounds in Sec.~\ref{sec:HallViscosityQGP::subsec::Alekseev3D}. The holographic framework thus provides a natural arena to assess the magnitude and scaling behavior of these coefficients at strong coupling, and to contrast them with the kinetic-theory estimates discussed earlier.

In Ref.~\cite{Ammon:2020rvg}, the two Hall viscosities were computed for a strongly coupled $\mathcal{N}=4$ Super-Yang-Mills theory at a large number of colors, subject to a strong external magnetic field, at finite temperature and chemical potential. 
For small values of dimensionless magnetic field $\tilde{B} := B/T^2$, the following behavior was found for $\tilde{\eta}_\parallel$
\begin{equation}
    \tilde{\eta}_\parallel \propto \tilde{B}^3 T^3 \, ,
\end{equation}
where $T$ is the temperature. Firstly, this differs from the corresponding behavior in kinetic theory \eqref{eq:visc-B}. If we expand $\tilde{\eta}_\parallel$ of \eqref{eq:visc-B} near $\freq \rightarrow 0$, we obtain $\tilde{\eta}_\parallel = \eta_0 \left( \freq - \freq^3/4 + \mathcal{O}(\freq^5) \right)$. So, at a small magnetic field, the leading behavior is $ \tilde{\eta}_\parallel \propto \tilde{B} T^3$ in kinetic theory, whereas in holography, it is $\tilde{\eta}_\parallel \propto \tilde{B}^3 T^3$.

The results from kinetic theory and holography also differ at large magnetic fields. In particular, \cite{Ammon:2020rvg} obtained following expression for $\tilde{\eta}_\parallel$ in the presence of large magnetic fields
\begin{equation}\label{eq:eta-par-holo}
    \tilde{\eta}_\parallel \approx 0.305 \tilde{\mu} \tilde{B}  T^3\, ,
\end{equation}
where $\tilde{\mu} := \mu/T$ is the dimensionless chemical potential. The kinetic theory expression for $\tilde{\eta}_\parallel$ \eqref{eq:visc-B}, when expanded around $\freq \rightarrow \infty$ gives $\tilde{\eta}_\parallel = \eta_0 \left( 4/\freq - 16/\freq^3 + \mathcal{O}(\freq^{-5}) \right)$. Thus, kinetic theory predicts $\tilde{\eta}_\parallel \propto T^3/\tilde{B}$ at large magnetic fields in contrast to the behavior $\tilde{\eta}_\parallel \propto \tilde{B} T^3$ obtained in holography.

We take $T \approx 300 $ MeV and $B \approx 10^{19}$ Gauss $= 0.195 \;  \text{GeV}^2$, i.e. $B \approx 2.17 \;  T^2$. From Fig.~3 of Ref.~\cite{Dore:2020jye}, for initial energy density of $\rho_0 = 1 fm^{-3} $ we can infer $\tilde{\mu} = \mu /T \approx 1$ assuming $\mu $ to be the baryon chemical potential. Substituting these in Eq.~\ref{eq:eta-par-holo}, we get 
\begin{align}
    \frac{\tilde{\eta}_\parallel}{T^3}  \approx 0.305 \tilde{\mu} \tilde{B}  
     = 0.305 \times \frac{\mu}{T} \times \frac{B}{T^2} 
     \approx 0.305 \times 1 \times \frac{2.17 \; T^2}{T^2} 
    = 0.66 \, .
\end{align}
Recalling that the zero magnetic field viscosity is $\eta_0 \approx 1.95 \, T^3$, we finally obtain $\tilde{\eta}_\parallel \approx 0.34  \eta_0$. 
This value is of the same order of magnitude as the corresponding kinetic-theory estimate for $\tilde{\eta}_\parallel$, Eq.~\eqref{eq:visc-B-ket}, suggesting qualitative consistency between weak- and strong-coupling approaches despite their differing functional dependence on the magnetic field.

For the transverse Hall viscosity, the holographic computation in~\cite{Ammon:2020rvg} found that $\tilde{\eta}_\perp = 0$ in this holographic model. 
This vanishing of $\tilde{\eta}_\perp$ is not enforced by symmetry but is instead a feature of the specific Einstein-Maxwell-Chern-Simons model considered in Ref.~\cite{Ammon:2020rvg}. 
This vanishing can be traced back to the fact that~\cite{Ammon:2020rvg} does not consider Chern-Simons terms coupling curvature to gauge fields in the gravitational action, which would lead to a gauge-gravitational anomaly in the dual field theory~\cite{Landsteiner:2016stv,Landsteiner:2016led}. This gauge-gravitational anomaly is included in the holographic model of~\cite{Landsteiner:2016stv} and leads to a non-vanishing $\tilde\eta_\perp$. More specifically, for the transverse Hall viscosity not to vanish, the retarded correlator $\langle T^{xz} (T^{xx}-T^{zz})\rangle$ needs to be non-zero, as seen from the Kubo formula~\eqref{eq:KuboFormulaeHallViscosities}.\footnote{These Kubo formulae were derived first specifically for Hall viscous transport in a theory with $U(1)_\mathrm{axial}\times U(1)_\mathrm{vector}$ containing a $U(1)^3$ anomaly along with an electromagnetic contribution to the axial anomaly as well as a mixed gauge-gravitational anomaly in~\cite{Landsteiner:2016stv}, and later including all transport effects for a theory with a single $U(1)^3$ anomaly in~\cite{Ammon:2020rvg}.} Non-vanishing of that correlator implies that the dual gravitational action couples the corresponding metric components to each other, i.e., the metric fluctuation $h_{xz}$ needs to couple to $(h_{xx}-h_{zz})$. This is achieved by the terms coupling curvature to the gauge fields, $\epsilon^{\mu\nu\rho\sigma\gamma} A_\mu R^\beta_{\delta \nu\rho} R^\delta_{\beta\sigma\gamma}$, with the Riemann tensor $R^\beta_{\delta \nu\rho}$ and the five-dimensional gauge field $A_\mu$. As mentioned above, such terms are dual to the gauge-gravitational anomaly. 
For the holographic model~\cite{Landsteiner:2016stv}, an analytic relation $\eta_\parallel/\eta_\perp=2\tilde\eta_\parallel/\tilde\eta_\perp$ holds between shear viscosities and Hall viscosities. Combining this relation with our holographic estimate  $\tilde\eta_\parallel\approx 0.34\eta_0$ yields $\tilde\eta_\perp=2\tilde\eta_\parallel{\eta_\perp}/{\eta_\parallel}\approx 0.68\eta_0$, when using that approximately $\eta_\perp/s\approx \eta_\parallel/s$ at $\tilde B\approx 2.17$ and $\tilde\mu\approx 1$, see figures 9 and 10 in~\cite{Ammon:2020rvg}. 

In summary, taken together, the holographic models~\cite{Landsteiner:2016stv} and~\cite{Ammon:2020rvg} imply that the axial $U(1)^3$ anomaly leads to a non-vanishing longitudinal Hall viscosity while the mixed gauge-gravitational anomaly leads to a non-vanishing of the transverse Hall viscosity. Both Hall viscosities at realistic QGP parameters are  of the same order  of magnitude as the standard shear viscosity $\eta_0$.

\section{Observables}\label{sec:HallObservables}
Since Hall viscosity is non-dissipative, it does not contribute to entropy production. Its effects, therefore, manifest not as damping but as systematic rotations and couplings between different shear components of the flow. 
We interpret the Kubo formulas,~\eqref{app:eq:kubo:eta-perp} and~\eqref{app:eq:kubo:eta-par} for the two Hall viscosities $\tilde{\eta}_\perp$ and $\tilde{\eta}_\parallel$ to understand how they influence the expanding (and rotating) QGP fireball. 
Let us begin with $\tilde{\eta}_\perp$.
For this, consider equations~\eqref{eq:const-3d:sub1},\eqref{eq:const-3d:sub2}, and~\eqref{eq:const-3d:sub3}. We observe that in the presence of non-zero $\tilde{\eta}_\perp$, shear in the reaction plane ($xz$-plane), $V_{xz}$, will source a pressure anisotropy ($P_{x} - P_{z}$) in the reaction plane and vice versa. Since the axis of rotation of the fireball is the $y$-axis, the rotation of the fireball induces shear in the $xz$-plane. Therefore, the effect of $\tilde{\eta}_\perp$ is to induce or enhance in-plane pressure anisotropy, $(P_{x} - P_{z})$. Let us understand why the rotation along the $y$-axis induces shear in the $xz$-plane. Note that the angular velocity can be expressed in terms of the velocity as $\omega = v/r$, where $r$ is the distance from the center of the fireball. Assuming that the entire QGP fireball rotates with a constant angular velocity $\omega$, the fluid layers farther from the center of the fireball move faster, i.e., have larger $v$, than the layers nearer the center of the fireball. This is equivalent to having shear in the $xz$-plane: the fluid elements moving along the $x$-axis move at different velocities at different $z$-values. Therefore, we observe that indeed the rotation of the fireball induces shear in the reaction plane. 

\begin{figure}[t]
\centering
\includegraphics[angle=-90,width=0.75\textwidth]{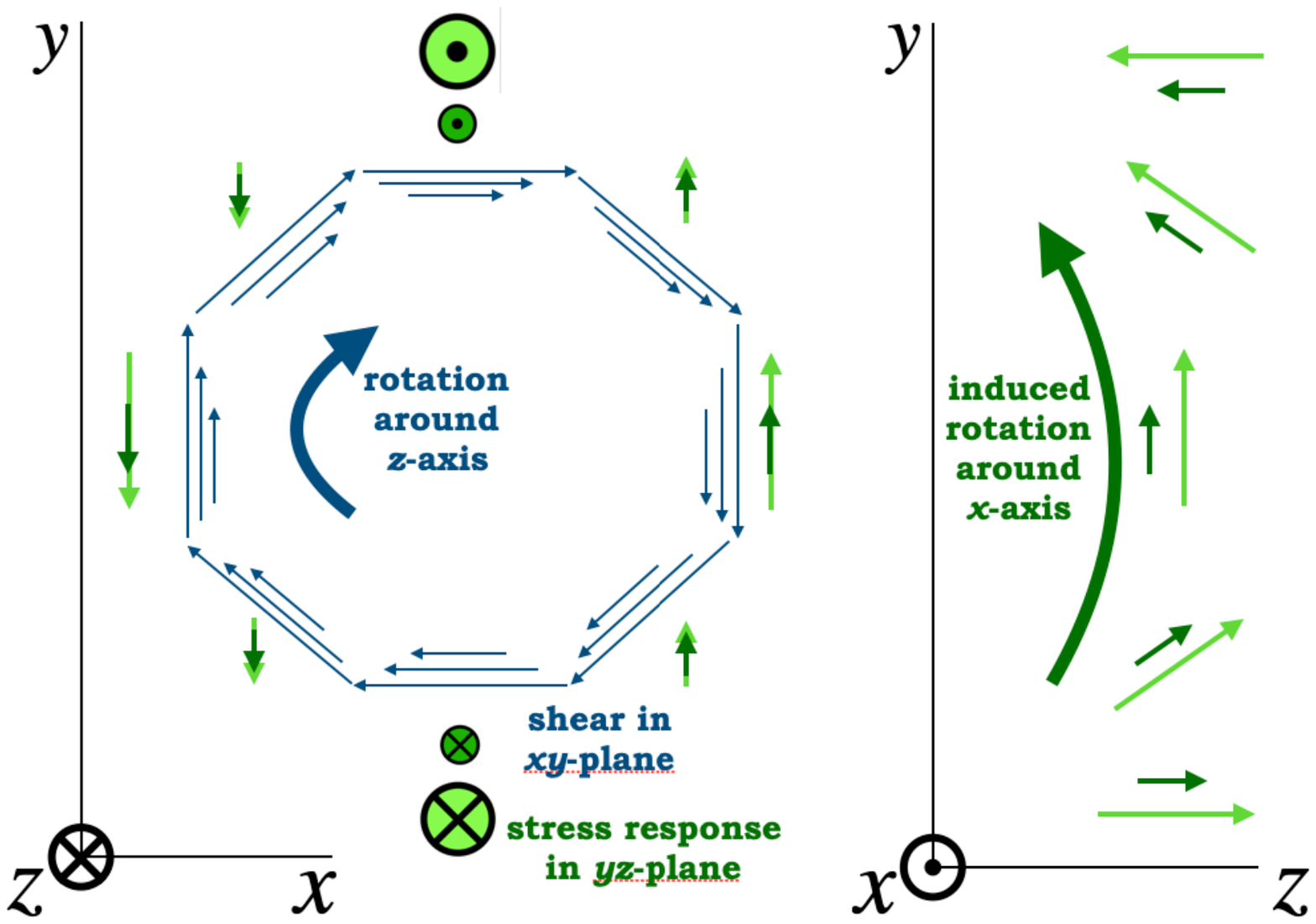}
\caption{\label{fig:RotationFromRotation}  
Illustration of the effect of $\tilde\eta_\parallel$: Rotation around the $z$-axis induces a rotation around the $x$-axis. {\it Left:} The shear flow in the $xy$-plane generated by rotation around the $z$-axis is schematically displayed by blue arrows. Longitudinal Hall viscosity induces a stress response in the $zy$-plane, $T^{zy}$, indicated by the green arrows. {\it Right:} The stress response (shear flow), $T^{zy}$, is displayed in the $zy$-plane by green arrows, and the big arrow indicates the induced rotation around the $x$-axis resulting from that shear flow.} 
\end{figure}

Next, to understand the effect of $\tilde{\eta}_\parallel$, we examine equations~\eqref{eq:const-3d:sub4} and \eqref{eq:const-3d:sub5}. They state that shear in the $zy$-plane induces shear in the $xy$-plane, and vice versa. As we have understood that a rotation about an axis induces shear in the plane perpendicular to the axis of rotation, we can interpret the effect of $\tilde{\eta}_\parallel$ as follows: rotations about the two axes in the reaction plane will source each other. In other words, a rotation about the $x$-axis will induce a rotation about the $z$-axis and vice versa, {c.f.} the illustration in Fig.~\ref{fig:RotationFromRotation}.

\subsection{Quantitative estimates of Hall viscosity effects on the QGP}
\hspace{0.5 cm} We aim to provide quantitative estimates for the terms in the constitutive relations~\eqref{eq:const-3d} that involve Hall viscosities, thereby assessing their impact on the QGP {stress-energy tensor}. These terms are evaluated at the proper time $\tau_0 = 1 \; \mathrm{fm}/c$, which corresponds to the typical initialization time for hydrodynamic simulations of the QGP. We will refer to $\tau_0$ as the hydrodynamic initialization time.
While the electromagnetic field produced by the spectator charges reaches its maximum strength
at very early times, it subsequently decays due to both geometric dilution and the finite electrical
conductivity of the medium. 
Our estimates correspond to early times relevant for the onset of hydrodynamic evolution
($\tau \sim \tau_0$), where the magnetic field, although already decaying from its initial peak, can still be sizable. In particular, several analyses that include the electromagnetic response of a
conducting QGP indicate that the magnetic field can remain non-negligible up to times of order
$\tau_0$, and in some scenarios even longer, depending on the assumed conductivity and
medium evolution~\cite{Shen:2025unr,Tuchin:2015oka,Stewart:2021mjz}. 
Anisotropic conductivities in the plasma subject to a strong magnetic field were recently discussed in~\cite{Ghosh:2024fkg,Ghosh:2024owm,Shovkovy:2025yvn}. {These results do not by themselves fix the field strength at
$\tau_0$.}

{Since we assumed that Hall viscosity is generated on the timescale $\tau_2 \approx 0.4 \, \mathrm{fm}/c$ given by holography \cite{Balasubramanian:2010ce, Muller:2020ziz}, 
its contribution to the viscous stress tensor does not require a long-lived magnetic field.
Instead, we can assume that Hall viscous stresses are generated whenever sizable shear gradients coexist with a
non-zero magnetic field, even if the latter is transient on a timescale longer than $\tau_2$. 
Consequently, the estimates presented here should be interpreted as capturing the Hall viscous stress corrections imprinted on the system at hydrodynamic initialization time $\tau_0>\tau_2$. 
These early-time corrections can subsequently influence the evolution of flow observables,
even if the magnetic field decays rapidly at later times.}

All terms on the RHS in the constitutive relations~\eqref{eq:const-3d} are expressed in terms of viscosities and accompanying stresses $V_{ij}$. Here, $V_{ij} := \partial_j V_i + \partial_i V_j$, where $ V_i = \langle v_i \rangle$ is the hydrodynamic flow velocity. We have already performed quantitative estimates of various viscosities that appear in the constitutive relations~\eqref{eq:const-3d} in the previous section. We now compute the values of $V_{ij}$. This can be readily done by adapting the ansatz of Ref.~\cite{Torrieri:2002jp} for the flow-velocity $u^\mu$ at the hydrodynamic initialization time $\tau_0$ given by
\begin{align}\label{eq:umu-giorgio}
     u^\mu = (\cosh y_L  \cosh y_T, \sinh y_T \cos \phi, \sinh y_T \sin \phi, \sinh y_L \cosh y_T )  \, ,
\end{align}
where $y_L(x,y)$ is the longitudinal momentum rapidity w.r.t. the beam (in polar coordinates 
\[\ \sinh y_L=\sinh y_T\sin\theta \] where $\theta$ is the longitudinal angle) and $y_T(x,y,z)$ is the transverse momentum rapidity and $\phi = \arctan \left( y/x \right)$. As given in Ref.\cite{Ryu:2021lnx},
\begin{align}\label{eq:yL}
    y_L(x,y) & = f\;  y_{CM}(x,y), \quad 0\leq f \leq 1,  \nonumber \\
    y_{CM}(x,y) &= \text{arctanh} \Big[ \frac{T_A - T_B}{T_A + T_B} \Big] \text{tanh}( y_{\text{beam}}),  \nonumber \\
    y_{beam} &= \text{arccosh} \left( \sqrt{s_{NN}}/ (2 m_N) \right) \, ,
\end{align}
where $y_{CM}(x,y)$ is the center-of-mass rapidity at the location $(x,y)$ in the transverse plane. $f \in [ 0, 1]$ is a parameter that controls the fraction of longitudinal momentum attributed to the flow velocity. $T_A(x,y)$, $T_B(x,y)$ are the participant thickness functions in the transverse plane, $m_N$ is the mass of the nucleon, $\sqrt{s_{NN}}$ is nucleon–nucleon center-of-mass energy, and $y_{\text{beam}}$ is the beam rapidity.

If the transverse flow is only associated with thermalization, we can set $y_T = 0$ at hydrodynamic initialization time $\tau_0$. However, initial transverse flow has been advocated to solve the so-called HBT puzzle~\cite{Pratt:2008bc}. In order to obtain analytic expressions for $y_T$, we use the ``\textit{universal flow}'' formula of Ref.~\cite{Vredevoogd:2008id}, which is based on the free-streaming limit. Accordingly the expression for $y_T$ is~\cite{Vredevoogd:2008id},
\begin{align}
    \tanh y_T =  - \frac{T^{tx}}{T^{tt}} \, .
\end{align}
Note that $u^\mu$ defined in \eqref{eq:umu-giorgio} is normalized $u^\mu u_\mu  = 1$ in Minkowski spacetime $g_{\mu \nu} = (1, -1, -1, -1)$.  Recall
\begin{align}
    u^\mu &= \gamma (c, v_x, v_y, v_z), \quad \gamma = \frac{1}{\sqrt{1 - \frac{v_x^2 + v_y^2 + v_z^2}{c^2}}}, \nonumber\\
     u^t &= \gamma c, \quad u^i = \gamma v_i, \;\; (i=x,y,z),  \nonumber \\
     v_i &= c \frac{u^i}{u^t} \, .
\end{align}
Comparing this with $u^\mu$ ansatz Eq.~\eqref{eq:umu-giorgio}, we get
\begin{subequations}
\begin{align}
    v_x &=c \tanh y_T \frac{\cos \phi}{ \cosh y_L} \, , \\
    v_y &=c \tanh y_T \frac{\sin \phi}{ \cosh y_L} \, , \\
    v_z &=c \tanh y_L \, .
\end{align}
\end{subequations}
To get $\tanh y_T$, we need expressions for $T^{tt}$ and $T^{tx}$. $T^{tt}$ can be expressed in terms of $T^{\tau \tau}$, $T^{\tau \eta}$, and $T^{\eta \eta}$ using the Milne-to-Minkowski coordinate transformation as 
\begin{align}\label{eq:T-Ml-Mn}
   T^{tt} & = (\cosh \eta)^2 T^{\tau \tau} + 2 \tau_0 \cosh \eta \sinh \eta T^{\tau \eta} + (\tau_0 \sinh \eta)^2 T^{\eta \eta}, \nonumber \\
          & = \left( 1 + \mathcal{O}(\eta)^2 \right) T^{\tau \tau} + 2 \tau_0  \left( \eta + \mathcal{O}(\eta)^3 \right) T^{\tau \eta} + \tau_0^2 \mathcal{O}(\eta)^2 T^{\eta \eta}   , \quad \eta \ll 1, \nonumber \\
          & = T^{\tau \tau} + 2 \tau_0 \eta T^{\tau \eta}  + \mathcal{O}(\eta)^2 , \quad \eta \ll 1 \, .
\end{align}
In the limit $\eta \ll 1$, we neglected $\mathcal{O}(\eta)^2$ terms and kept terms up to linear order in $\eta$. At the hydrodynamic initialization time $\tau_0$, $\eta \ll 1$ is a valid assumption. Expressions for $T^{\tau \tau}$ and $T^{\tau \eta}$ at hydrodynamic initialization time $\tau_0$ as given in Ref.~\cite{Ryu:2021lnx} are
\begin{align}
    T^{\tau \tau} (x, y, \eta) &= e(x, y, \eta) \text{cosh}(y_L) \label{eq:Ttautau} \, , \\
    T^{\tau \eta} (x, y, \eta) &= \frac{1}{\tau_0} e(x, y, \eta) \text{sinh}(y_L) \label{eq:Ttaueta} \, .
\end{align}
The expression for energy density $e(x,y,\eta)$ is given by~\cite{Ryu:2021lnx} ($\Theta(...)$ is the Heavyside function)
\begin{align}\label{eq:energy-RJS}
    e(x,y,\eta; y_{CM} - y_L) &= \mathcal{N}_e (x,y) \text{exp} \Bigg[ - \frac{\left( | \eta - (y_{CM} - y_L) | - \eta_0 \right)^2}{2 \sigma_\eta^2}  \Theta( | \eta - (y_{CM} - y_L) | - \eta_0)\Bigg], \nonumber \\
    \mathcal{N}_e (x,y) &= \frac{M(x,z)}{2 \text{sinh}(\eta_0) + \sqrt{\pi/2} \sigma_\eta e^{\sigma_\eta^2/2} C_\eta}, \nonumber \\
    C_\eta &= e^{\eta_0} \text{erfc}\left( -\sqrt{\frac{1}{2}} \sigma_\eta \right) + e^{-\eta_0} \text{erfc} \left( \sqrt{\frac{1}{2}} \sigma_\eta \right) \, ,
 \end{align}
where erfc$(x)$ is the error function. The parameter $\eta_0$ determines the width of the plateau and $\sigma_\eta$ controls how fast the energy density falls off at the edge of the plateau. Ref.~\cite{Ryu:2021lnx} and Ref.~\cite{Alzhrani:2022dpi} set $\eta_0 = \text{min} (\eta_0, y_{\text{beam}} - (y_{CM} - y_L))$.

The invariant mass $M(x,y) $ can be expressed in terms of the participant thickness functions as follows,
\begin{align}\label{eq:Mxy}
    M(x,y) = m_N \sqrt{ T_A^2 + T_B^2 + 2 T_A T_B \text{cosh}(2 y_{\text{beam}})} \, .
\end{align} 
Since we are only interested up to linear order in small $\eta$, we fix a small bound $ |\eta | \leq \varepsilon $ with $\varepsilon \ll \eta_0$. Note that Table I in Ref.\cite{Alzhrani:2022dpi} gives $\eta_0 = 2.5$, so this assumption is reasonable.  If the transverse region satisfies 
\begin{equation}
    |  y_{CM}(x,y) - y_L (x,y) | = | (1-f)y_{CM}(x,y)| \leq \eta_0 - \varepsilon, \quad \forall x,y \, ,
\end{equation}
then, by triangle inequality, one obtains
\begin{equation}
   |\eta - (y_{CM}(x,y) - y_L (x,y) )|=  |\eta - (1-f)y_{CM}(x,y)| \leq |\eta| + |(1-f)y_{CM}(x,y)| \leq \varepsilon + \eta_0 - \varepsilon = \eta_0 \, .
\end{equation}
This implies 
\begin{equation}
    \Theta( |\eta - (y_{CM}(x,y) - y_L (x,y) )| - \eta_0) = 0 \, .
\end{equation}
Therefore, we can drop the exponential term in the energy density expression and get
\begin{equation}\label{eq:e=Ne}
    e(x,y) = \mathcal{N}_e (x,y)  \, .
\end{equation}
To obtain $T^{tx}$, we use the universal flow formula of Ref.~\cite{Vredevoogd:2008id}
\begin{align}\label{eq:univ-flow-Pratt}
    \frac{T^{tx}}{T^{tt}} &\approx \frac{t}{2} \frac{\partial_x T^{tt}}{ T^{tt}}  
    =  \frac{\tau_0 \cosh \eta}{2} \frac{\partial_x T^{tt}}{ T^{tt}} \, ,
\end{align}
where in the second equality we used $t = \tau_0 \cosh \eta$; $t$ is Minkowski time and $\tau_0$ is the hydrodynamic initialization proper time.

The gradients of velocities $v_x$, $v_y$, and $v_z$ are
\begin{subequations}\label{eq:vel-grad}
\begin{align}
    \partial_x v_x &= c \frac{\cosh y_L \cos \phi \partial_x \tanh y_T - \cosh y_L \tanh y_T  \sin \phi  \partial_x \phi - \tanh y_T \cos \phi  \sinh y_L \partial_x y_L}{\cosh^2 y_L} \, , \\
    \partial_y v_x &= c \frac{\cosh y_L \cos \phi \partial_y \tanh y_T - \cosh y_L \tanh y_T  \sin \phi \partial_y \phi- \tanh y_T \cos \phi \sinh y_L \partial_y y_L}{\cosh^2 y_L}  \, , \\
    \partial_z v_x &= c \frac{\cosh y_L \cos \phi \partial_z \tanh y_T}{\cosh^2 y_L} \, , \\
    \partial_x v_y &= c \frac{\cosh y_L \sin \phi \partial_x \tanh y_T + \cosh y_L \tanh y_T  \cos \phi \partial_x \phi - \tanh y_T \sin \phi \sinh y_L \partial_x y_L}{\cosh^2 y_L} \, , \\
    \partial_y v_y &= c \frac{\cosh y_L \sin \phi \partial_y \tanh y_T + \cosh y_L \tanh y_T \cos \phi \partial_y \phi - \tanh y_T \sin \phi \sinh y_L \partial_y y_L}{\cosh^2 y_L} \, , \\
    \partial_z v_y &= c \frac{\cosh y_L \sin \phi \partial_z \tanh y_T }{\cosh^2 y_L} \, , \\
     \partial_x v_z &= c \; \text{sech}^2 y_L\partial_x y_L \, , \\
    \partial_y v_z &= c \; \text{sech}^2 y_L \partial_y y_L \, , \\
    \partial_z v_z &= 0 \, .
\end{align}
\end{subequations}
We need gradients of $\phi$, $y_L$, and $\tanh y_T$ to evaluate the velocity gradients above. The expressions for these gradients can be found in the Appendix~\ref{app:velocity-grad}.

We need to choose some differentiable functional form of participant thickness functions $T_A(x,y)$ and $T_B(x,y)$. The participant thickness functions are defined by
\begin{equation}\label{eq:TA-defn}
    T_{A(B)}(x,y) = \int_{-\infty}^{\infty} d z \rho(\sqrt{x^2 + y^2 + z^2 }) \, ,
\end{equation}
where $\rho(\mathfrak{r})$, $(\mathfrak{r} = \sqrt{x^2 + y^2 + z^2 })$ is the nuclear density function. The most common choice for $\rho(\mathfrak{r})$ is the Wood-Saxon density given by
\begin{equation}\label{eq:Wood-Saxon}
    \rho(\mathfrak{r}) = \frac{\rho_0}{1 - \exp \left( \frac{\mathfrak{r} - R_0}{a} \right)} \, .
\end{equation}
However the integral Eq.~\eqref{eq:TA-defn} cannot be computed analytically for Wood-Saxon density Eq.~\eqref{eq:Wood-Saxon}.

Since we are only interested in a back-of-the-envelope estimate, we choose the simplest ansatz for the nuclear density function - that of a ``hard sphere'' nucleus: uniform nuclear density with a sharp edge 
\begin{equation}\label{eq:hard-sph}
    \rho(\mathfrak{r}) = \rho_0 \Theta(R - \mathfrak{r}) \, ,
\end{equation}
where $R$ is the nuclear radius and $\Theta$ is the Heavyside theta function. We choose values of parameters $\rho_0$ and $R$ in the hard sphere nuclear density function $\rho(\mathfrak{r})$ of Eq.~\eqref{eq:hard-sph} as follows. 
For a given nucleus with mass number $A$,
  \begin{equation}
      A = \int d^3 \mathfrak{r} \rho(\mathfrak{r}) = \frac{4}{3} \pi R^3 \rho_0 \, .
  \end{equation}
We fix $\rho_0$ to be the nuclear saturation density and solve for $R$. The nuclear saturation density is $ 0.16$ fm$^{-3}$~\cite{PhysRevC.102.044321, Drischler:2024ebw, hjn1-24xg, Zou:2023quo}. We consider Au nuclei which have $A=197$. This gives $R = 6.65 fm$.

By integrating $\rho(\mathfrak{r})$ over $z$, we get
\begin{equation}
    T_{A(B)}(x,y) = 
    \begin{cases}
2 \rho_0 \sqrt{R^2 - (\mathfrak{r}^{A(B)}_\perp)^2}, & \mathfrak{r}^{A(B)}_\perp \leq R\\
0, & \mathfrak{r}^{A(B)}_\perp > R 
\end{cases} \, ,
\end{equation}
where 
\begin{align}
    \mathfrak{r}^{A}_\perp (x,y;b) &= \sqrt{\left( x - \frac{b}{2}\right)^2 + y^2} \, , \\
    \mathfrak{r}^{B}_\perp (x,y;b) &= \sqrt{\left( x + \frac{b}{2}\right)^2 + y^2}  \, ,
\end{align}
with the impact parameter is $b$, and we choose centers of nucleus A and nucleus B at transverse positions $(x,y) = (\frac{b}{2},0)$ and $(x,y) = (-\frac{b}{2},0)$, respectively. The values of velocity gradients of Eq.~\eqref{eq:vel-grad} are provided in Table~\ref{tab:vel-grad}. The values of various terms appearing in constitutive relations Eq.~\eqref{eq:const-3d} are provided in Tables~\ref{tab:stresses-1} and~\ref{tab:stresses-2}. From Tables~\ref{tab:stresses-1} and~\ref{tab:stresses-2}, we observe that the terms with Hall viscosities are comparable to the terms with anisotropic shear viscosities.

\begin{table}[h!]
\centering
\begin{tabular}{|c|c|c|c|c|c|c|c|c|c|}
\hline
\textbf{(x,y,z)}  & \textbf{$\partial_x v_x$}  & \textbf{$\partial_y v_x$} & \textbf{$\partial_z v_x$}  & \textbf{$\partial_x v_y$}  & \textbf{$\partial_y v_y$}  & \textbf{$\partial_z v_y$} & \textbf{$\partial_x v_z$}  & \textbf{$\partial_y v_z$} & \textbf{$\partial_z v_z$}  \\ 
\hline\hline
(0.001,0,0) & 0.05569 & 0 & -0.02815 & 0 & 0.05569 & 0 & 0.02815 & 0 & 0 \\ 
\hline
(0.5,0.5,0.1) & 0.06577 & -0.017653 & -0.01953 & 0.02758 & 0.020543 & -0.019537 & 0.03048 & 0.0006358 & 0 \\ 
\hline
(1,1,0.1) & 0.1108 & -0.01613 & -0.02292 & 0.05889 & 0.035781 & -0.022925 & 0.04023 & 0.003356 & 0 \\ 
\hline
\end{tabular}
\caption{Note that $\eta = \operatorname{arcsinh}(z/\tau_0)$. For $z = 0.1 fm$, we get $\eta = 0.0998 $ for $\tau_0 = 1 fm/c$. Since our expressions for velocity gradients are valid only up to linear order in $\eta$ around $\eta = 0$, we do estimates for $z \leq 0.1 fm$. The values of parameters $f$, $\eta_0$, $\sigma_\eta$, $\tau_0$ are taken from table I of Ref.~\cite{Alzhrani:2022dpi}, namely $f=0.15$, $\eta_0 = 2.5$, $\sigma_\eta = 0.5$, $\tau_0 = 1 \; fm/c$. $\sqrt{s_{NN}} = 200$ GeV, $m_N = 0.94$ GeV$/c^2$. $\rho_0$ is taken as nuclear saturation density, $0.16 fm^{-3}$ and $A = 197$ (Au mass number). From $\rho_0$ and $A$, the nuclear radius $R$ is determined by using $A = \rho_0 \frac{4}{3} \pi R^3$ to be $R = 6.65 fm$. The value of the impact parameter, $b$, is taken as $b = 9 fm$ (40-50 $\%$ centrality class ). Note that in the table, the positions $(x,y,z)$ are in units of $fm$ and velocity gradients are in units of $fm^{-1}$. We have used natural units, i.e., $\hbar =c =k_B= 1$.}
\label{tab:vel-grad}
\end{table}

\begin{table}[h!]
\centering
\begin{tabular}{|c|c|c|c|}
\hline
\textbf{(x,y,z)} & $\frac{\eta_0}{2} (V_{xx} + V_{zz})$ &  $\frac{\eta_\perp}{2} (V_{xx} - V_{zz})$ & $ \tilde{\eta}_\perp V_{xz}$  \\ 
\hline\hline
(0.001,0,0) & 0.000578 & 0.000537 & 0 \\
\hline
(0.5,0.5,0.1) & 0.000683 & 0.000635 & 0.0000284 \\
\hline
(1,1,0.1) & 0.00115 & 0.00107 & 0.0000449 \\
\hline
\end{tabular}
\caption{In this table and in Table~\ref{tab:stresses-2}, we estimate values of various terms appearing in constitutive relations, Eq.~\eqref{eq:const-3d}, at various spacetime locations $(x,y,z)$. We have used the velocity gradients of Table~\ref{tab:vel-grad}. Note that the values of velocity gradients in Table~\ref{tab:vel-grad} are in units of $fm^{-1}$ (in natural units $\hbar = c =1$). We used the relation $1 fm^{-1} = 0.1973 $ GeV to convert them in units of GeV. Furthermore we used the zero-magnetic field viscosity $\eta_0 $ values from our earlier estimate, namely, $\eta_0 = 0.0526 \; GeV^3$. Values of the other four viscosities were also used from our earlier estimate of Eq.~\eqref{eq:visc-B-ket}, namely {$\eta_\perp =0.93 \eta_0$, $\eta_\parallel = 0.98 \eta_0$, $\tilde{\eta}_\perp = 0.25 \eta_0$, $\tilde{\eta}_\parallel = 0.27 \eta_0$}. All values in this table are in the units of $GeV^4$. Estimates of three of the terms from constitutive relations, Eq.~\eqref{eq:const-3d}, are in this table, whereas the estimates of the remaining six terms from constitutive relations, Eq.~\eqref{eq:const-3d}, are in Table~\ref{tab:stresses-2}. }
\label{tab:stresses-1}
\end{table}

\begin{table}[h!]
\centering
\begin{tabular}{|c|c|c|c|c|c|c|c|}
\hline
\textbf{(x,y,z)} & \textbf{$\eta_\perp  V_{xz}$} & \textbf{$\frac{\tilde{\eta}_\perp}{2}  \left( V_{xx} - V_{zz}\right)$} & \textbf{$ \eta_\parallel V_{xy}$} & \textbf{$ \frac{\tilde{\eta}_\parallel}{2} V_{zy}$} & \textbf{$ \eta_\parallel V_{zy}$} & \textbf{$ \frac{\tilde{\eta}_\parallel}{2} V_{xy} $}   \\
\hline\hline
(0.001,0,0) & 0 & 0.000144 & 0 & 0 & 0 & 0 \\
\hline
(0.5,0.5,0.1) & 0.000106 & 0.000171 & 0.000101
& -0.0000265 & -0.000192 & 0.0000139 \\
\hline
(1,1,0.1) & 0.000167 & 0.000287 & 0.000435
& -0.0000274 & -0.000199 & 0.0000599 \\
\hline
\end{tabular}
\caption{Together in this table and in Table~\ref{tab:stresses-1}, we estimate the values of the various terms appearing in constitutive relations, Eq.~\eqref{eq:const-3d}, at various spacetime locations $(x,y,z)$. See caption of Table~\ref{tab:stresses-1}. }
\label{tab:stresses-2}
\end{table}

\subsection{Observables and their quantitative estimates}
Hydrodynamics is a deterministic theory. Once initial conditions are known, the dynamics, given by conservation laws
\begin{equation}
    \partial_\mu T^{\mu \nu}=\partial_\mu J^\nu=0
\end{equation}
as well as the EoS and transport coefficients (determining the form of $T_{\mu \nu},J_\mu$)
uniquely determines subsequent evolution up to initial state fluctuations (described by geometric models or approaches like the Color Glass) and final state fluctuations (assumed to be thermal as well as given by resonance decays).

In heavy-ion collisions (HIC) it is convenient to parametrize the details of the evolution in terms of a harmonic analysis w.r.t. the azimuthal angle, which is given by the impact parameter.
In other words, the anisotropy $\epsilon_u$ of the collective flow\footnote{For simplicity we set the longitudinal coordinate to zero, which for Bjorken expansion corresponds to mid-rapidity}
\begin{equation}
\label{uphi}
u_{r,\theta}=u_0(r) \left(1+\sum_n \epsilon_{un} \cos(n(\phi - \phi_{un}))\right)    \, , 
\end{equation}
which freezes out into particles, via a Cooper-Frye type ansatz, in terms of $u_\mu$, temperature $T$, and a freeze-out hypersurface $\Sigma_\mu$
\begin{equation}
E    \frac{dN}{d^3 p} = \int d \Sigma_\mu p^\mu \exp \left[ \frac{p_\mu u^\mu(x)}{T(x)} \right]  \, .
\end{equation}

Obvious observables to consider in this regard are the anisotropic flow coefficients $v_n$. At mid-rapidity these coefficients are defined by 
\begin{equation}
\label{vndef}
    \frac{dN}{p_T dp_T d\phi}= 1+\sum_n 2 v_n(p_T) \cos\left(n(\phi-\phi_{0n})\right) \, , \qquad  v_n \equiv \ave{\vn{n}} \, ,
    \end{equation}
and $v_n$ can be measured with a variety of techniques, including numerical Fourier decomposition, cumulants of $\phi$ distributions and Lee-Young zeroes (see references in~\cite{cmsexpv2}). 

In hydrodynamics, the $v_n$ are uniquely determined by $\epsilon_x$, the initial position space anisotropy of energy density $e$ (given by azimuthal gradients),
\begin{equation}
\label{posphi}
    e(r,\theta,t=0)=e_0(r) \left(1+\sum_n \epsilon_{xn} \cos(n(\phi - \phi_{en}))\right) \, ,
    \end{equation}
as well as intensive parameters such as pressure and viscosity (which in turn depend on energy density). 

Of course hydrodynamic equations are highly non-linear, so mixings between different phases $\phi_0$, and flow and freeze-out Fourier coefficients $\epsilon_{un}$, and correlations between $\epsilon_{xn}$ and $v_{n\ne m}$ are non-negligible.
However, unless one is in a turbulent regime, a good control over initial conditions and a stable enough hydrodynamic code should lead to deterministic $v_n$ once initial geometry was precisely enough accounted for.

From the previous sections, it is straightforward to see that Hall viscosity would result in an event plane and tilt plane rotations with respect to their ''geometric'' expecations. 
In particular, the longitudinal component of vorticity is directly connected to the well-developed program of the study of fluctuations of anisotropic flow/event eccentricity.\footnote{The transverse component of vorticity, which dominates at lower energy, could also lead to interesting effects in rapidity, but the background of these fluctuations is not as well understood.}  We shall briefly discuss experimental observables sensitive to this, as well as make a few qualitative estimates.

Dynamical fluctuations can be constrained by $v_2$ fluctuation and correlations, expressible  event-by-event via symmetric cumulants\cite{atlas1,atlas2,star1,alice1,ante1} ($\Delta A=A-\ave{A}$)
\begin{equation}
\label{vncorr}
    \ave{\left( \delta v_n (y) \delta v_m(y') \right)} \equiv \langle \Delta \cos[n(\phi - \phi_{0n}(y))] \Delta \cos[m(\phi - \phi_{0m}(y')] \rangle_{event}
\end{equation}
\[  \sim  \ave{\Delta \cos\left(n \left( \sum_i^M \phi_i-\sum_j^M \phi_j\right)\right)  \Delta \cos\left(m \left( \sum_i^K \phi_k-\sum_j^K \phi_l\right)\right) }_{event}  \]
With a separation between y and y' \cite{ante2}. 
Higher order cumulants $M,K\gg 2$ and greater rapidity separation eliminats ''non-flow'' effects (jet fragmentation, resonance decays, momentum conservation and so on).     Hall viscosity is unusual in that it is a ''flow'' effect (and hence will persist for higher symmetric cumulants) but the rotation of $\phi_0$ will be local in rapidity.   Provided the detector has enough rapidity coverage and enough statistics to check cumulant scaling, the effect of Hall viscosity might qualitatively appear.

Correlations between $\phi_n$ and $\phi_m$ can also be isolated (if $y\rightarrow y'$ the cumulant is by construction independent of $\phi_0$).   More direct observables of $\phi_0$ are also possible \cite{evcorralice,evcorratlas} and, provided they can be done at wide rapidity separations (or an event plane detector is available) , they can directly infer a rotation of the azimuthal momentum event plane w.r.t. its geometric expectation.

Events can even be binned by 
$\cos[n (\phi - \phi_{0n})]$ leading to 
``event engineering'' \cite{engineering} (geometry selection) techniques.  So far no evidence of fluctuations due to dynamics have been found, see~\cite{vogel} and references therein. 

While a quantitative estimate of the effects sketched in Figs.~\ref{figvort} and~\ref{figvortlong} would need a numerical hydrodynamic calculation, it is possible to give an order-of-magnitude estimate of the effect from available experimental data and parameters characterizing the collision.  The previous section makes it clear that the effect we are looking for is a torque density, a torque divided by the relevant cross-sectional area.
Thus looking at Figs. \ref{figvort} and \ref{figvortlong} we get 
\begin{equation}
\frac{\torque_{\theta}}{(t_f-t_0)S_{xz}} \sim \tilde\eta_{\perp} (\Pi_{xx}-\Pi_{yy}) \, , \qquad  \frac{\torque_{\phi}}{(t_f-t_0)S_{yz}} \sim \tilde\eta_{\parallel} (
\Pi_{zz}-\Pi_{xx}) \,  ,
\end{equation}
where $\torque$ is the torque, $S$ a cross-sectional area and $\theta,\phi$ are, respectively, the angle between $x-z,x-y$. 
{It is important to emphasize that the Hall viscous effects discussed here are cumulative over the lifetime of the fireball. While the instantaneous Hall viscous stresses are small, their integrated effect over several fm/c can lead to observable phase shifts (i.e. shifts of the angles $\theta$ and $\phi$ inferrred from data w.r.t. the event geometry definitions of \eqn{eq:umu-giorgio})}
The resulting observable angles will then simply be the torque multiplied by the lifetime of the fireball squared, {i.e.}, the square of the difference between freezeout time $t_f$ and thermalization time $t_0$.
\begin{equation}
  \Delta \theta \sim \frac{1}{2} (t_f-t_0)^{2} \torque_\theta \,  , \qquad 
\Delta \phi \sim \frac{1}{2} (t_f-t_0)^{2} 
\torque_\phi\, .
\end{equation}
For an order of magnitude estimate, we note that the measured anisotropic flow coefficients directly track the shear,
\begin{equation}
v_2 \sim \frac{\Pi_{xx}-\Pi_{yy}}{\Pi_{xx}+\Pi_{yy}} \,  , \qquad \Delta y \frac{d v_1}{dy}\sim  \frac{\Pi_{zz}-\Pi_{xx}}{\Pi_{zz}+\Pi_{xx}} \, ,
\end{equation}
while the sum of the shears can be related to the average transverse momentum $\ave{p_T}$ together with the relevant volume of the fireball
\begin{equation}
\Pi_{xx}+\Pi_{yy} \sim   \frac{\ave{p_T}}{t_0 \Delta y } \, , \qquad  \Pi_{xx}+\Pi_{zz}
\sim   \frac{\ave{p}}{ t_0 \Delta y} 
\end{equation}
at mid-rapidity the longitudinal gradient is small so $\ave{p}\simeq \ave{p_T}$

Finally, the transverse areas are related to the eccentricity $\epsilon$ and the nuclear radius $R$ via elementary geometry
\begin{equation}
S_{xy} \sim \epsilon \, R^{2}  \, , \qquad S_{xz} \sim R \, t_0 \, \Delta y \, ,
\end{equation}
Thus, the observable phase shifts are given as a function of $v_1,v_2,\epsilon,R,\ave{p_T},\Delta y,t_0,t_f$ as
\begin{equation}
\label{quantresulttheta}
\Delta \theta
 \sim \frac{\left(t_f-t_0\right)}{ 2 \Delta y t_0}  \, \frac{ v_2  \,  R \, t_0 \,    }{\Delta y \ave{p_T}} \,  {\tilde{\eta}_\perp}  \, .
\end{equation}
\begin{equation}
\label{quantresultphi}
 \Delta \phi \sim \frac{\left(t_f-t_0\right)}{ 2 \Delta y t_0}  \, 
\frac{dv_1}{dy} \frac{\Delta y  
 \epsilon \, R^{2}  \,   }{ \ave{p_T} } \, 
{\tilde{\eta}_\parallel}  \, .
\end{equation}
This is essentially a back-of-the-envelope dimensional analysis estimate, but it gives a physical indication of what Hall viscosity actually does, as also summarized in the figures:  It causes an interplay between the transverse and longitudinal Fourier components and the longitudinal and transverse mean ``phase'' ($\phi_{0n}$ in \eqn{vndef}) . 

For the numerical estimate, 
we took $\ave{p_T}$ from \cite{avept} to be $\simeq 0.6$ GeV,
$p=p_T\sinh y$,
$t_0=1$ and $t_f=5$ fm, $R=6$fm \cite{glauber}, $\epsilon=0.2-0.8$ \cite{glauber}, $\Delta v_1/\Delta y =0.5-1$ \cite{starv1,starv1s}, $\Delta y = 1$ and $v_2=0.01-0.05$ \cite{starv2}. 
the numerical values for $\Delta \theta,\Delta\phi$ would be
\begin{equation}\label{eq:phase-shift-estimates}
\Delta \theta= \alpha_\perp    \sim GeV^{-3}
\left({15.41} \, \tilde{\eta}_\perp \right) \, , 
\qquad  \Delta \phi =\alpha_\parallel \sim GeV^{-3} \\
\left({1156} \,  \tilde{\eta}_\parallel \right)
\end{equation}
where 
\begin{equation} \alpha_i=\frac{\tilde{\eta}_i}{s} \, , \qquad s \sim \frac{4}{N_{part} fm^3}\frac{dN}{dy}\sim s_0 s_{cm}^{0.155} \ln s_{cm} \, .
\end{equation}
Entropy is estimated by assigning each particle 4 units of entropy~\cite{fermi}, and by estimating the multiplicity per participant according to an empirical formula in terms of the center of mass energy $s_{cm}$~\cite{multestimate}.

$\Delta \theta$ and $\Delta \phi$ are in principle observable, as they appear as differences between the phase $\phi_n$ in \eqn{posphi} and \eqn{uphi}.   \eqn{posphi} is related to the spectator recoil angle, which is observable with a zero degree calorimeter.  \eqn{uphi} relates to $\theta_0$ and $\phi_0$ phase angles that maximize $\ave{\cos(2(\phi-\phi_0))}$ and $\ave{\cos(\theta-\theta_0)}$ once non-flow has been taken out.   A systematic shift between them, related to $v_{1,2}$, and global event characteristics such as the nuclear radius and $\ave{p_T}$, would give an experimental indication of the presence of Hall viscosity. 
{Experimentally, these effects can be accessed through event-plane correlations and directed-flow rapidity slopes, all of which are currently measured by experiments such as STAR \cite{directed} and ALICE. Event-engineering techniques provide a particularly promising avenue for isolating Hall viscous signatures.}

{

\subsection{Sensitivity to magnetic-field strength}
\label{subsec:B-sensitivity}

\noindent\hspace*{0.4cm} The magnetic field can decay substantially before hydrodynamics
becomes applicable. Its magnitude during the hydrodynamic stage
depends on the electromagnetic response and evolution of the
medium. We therefore supplement the strong-field value used
above with a scan over a range of field strengths. We use the
Au+Au, $\sqrt{s_{NN}}=200$ GeV, $40$--$60\%$ centrality curve in
Fig.~3(b) of Ref.~\cite{Ayala2020MagneticField}, reproduced in
Fig.~4.5(b) of Ref.~\cite{Adhikari:2024bfa}. This UrQMD-based calculation includes
participant and spectator contributions, evaluates the field at
the collision centre, and omits electromagnetic medium response.
The five field strengths in Table~\ref{tab:B-sensitivity} are
approximate values sampled at times $t_B=0.1$--$0.5$ fm/$c$.

We use the thermal-mass kinetic-theory prescription of
Sec.~\ref{sec:HallViscosityQGP::subsec:OOM-estimate-KT}, retaining
$T=0.300$ GeV, $m=gT=0.582$ GeV,
and
$\eta_0=0.0526$ GeV$^3$. The viscosities follow from
Eq.~\eqref{eq:visc-B}, with
$\varpi=2|q_uB|\tau_2/m$ and $|q_uB|=2|eB|/3$.
The angular estimates use Eqs.~\eqref{quantresulttheta} and \eqref{quantresultphi}, with values of all parameters as given above Eq.~\eqref{eq:phase-shift-estimates}, except $t_0 = t_B$.

\begin{table}[h!]
\centering
\begin{tabular}{|c|c|c|c|c|c|c|c|}
\hline
$t_B$ & $|B|$
& $\eta_\perp/\eta_0$ & $\eta_\parallel/\eta_0$
& $\tilde{\eta}_\perp/\eta_0$
& $\tilde{\eta}_\parallel/\eta_0$
& $\Delta\theta$ & $\Delta\phi$ \\
$(\mathrm{fm}/c)$ & $(\mathrm{G})$
& & & & & $(\mathrm{rad})$ & $(\mathrm{rad})$ \\
\hline\hline
$0.1$ & $1.5\times10^{19}$ & $0.87$ & $0.96$
& $0.34$ & $0.38$
& $3.4\times10^{-1}$ & $2.8\times10^{2}$ \\
\hline
$0.2$ & $1.7\times10^{18}$ & $1.0$ & $1.0$
& $4.5\times10^{-2}$ & $4.5\times10^{-2}$
& $4.4\times10^{-2}$ & $1.7\times10^{1}$ \\
\hline
$0.3$ & $4.3\times10^{17}$ & $1.0$ & $1.0$
& $1.2\times10^{-2}$ & $1.2\times10^{-2}$
& $1.1\times10^{-2}$ & $2.8$ \\
\hline
$0.4$ & $2.2\times10^{17}$ & $1.0$ & $1.0$
& $6.1\times10^{-3}$ & $6.1\times10^{-3}$
& $5.7\times10^{-3}$ & $1.1$ \\
\hline
$0.5$ & $1.6\times10^{17}$ & $1.0$ & $1.0$
& $4.5\times10^{-3}$ & $4.5\times10^{-3}$
& $4.1\times10^{-3}$ & $6.1\times10^{-1}$ \\
\hline
\end{tabular}
\caption{{Sensitivity of the viscosities and
angular-shift estimates to the magnetic-field strength.}}
\label{tab:B-sensitivity}
\end{table}

Over the sampled range, decreasing the field suppresses both
Hall viscosities and the corresponding angular estimates, while
the dissipative viscosities approach $\eta_0$. At the strongest
field, the Hall viscosities are approximately
$0.34\eta_0$ and $0.38\eta_0$; at the weakest, both are approximately
$4.5\times10^{-3}\eta_0$. 

Looking at Table \ref{tab:B-sensitivity}, it is clear that if Hall viscosity is of the order of magnitude of other viscosities, and if magnetic fields in general are relevant to dynamics\footnote{This is not such a straight-forward assumption: If relaxation times are too long w.r.t. the lifetime of the fireball, as argued in \cite{xumagnetic}, or if the initial magnetic field is too disoriented, as possible from \cite{PhysRevC.85.044907,Bzdak:2011yy}, it might be that bulk dynamics remains insensitive to magnetic fields}, then Hall viscosity is amply observable.  In fact, combinations within  Table \ref{tab:B-sensitivity} where $\Delta \phi \gg 1$ are ruled out since they would mean the reference angle for elliptic flow is totally unrelated to the event plane, which is in contradiction to experimental data \cite{starv2}.   Thus, these back-of-the-envelope estimates motivate more detailed simulations and an experimental search.
}

\section{Conclusions and Outlook}\label{sec:ConcOutlook}

\hspace{0.5cm} In this work we have provided the first quantitative estimate of the Hall viscosity in the quark-gluon plasma (QGP), demonstrating that non-dissipative parity-odd transport can appear naturally once rotational symmetry is broken by the strong magnetic field 
in off-central heavy-ion collisions.

By extending the {Alekseev mechanism} from two-dimensional kinetic theory~\cite{alekseev2016negative} to three spatial dimensions, we have shown that two independent Hall viscosities emerge: a {transverse} ($\etprpt$) and a {longitudinal} ($\etpart$) component. These enter the viscous stress tensor through well-defined constitutive relations, Eq.~\eqref{eq:const-3d}, and modify the hydrodynamic response of the plasma in distinct ways: As illustrated in Fig.~\ref{fig:HallViscosities}, $\etprpt$ couples in-reaction plane shear to pressure anisotropy in the reaction plane, while $\etpart$ couples mixed in- and out-of-reaction plane shears with each other.

Order-of-magnitude estimates based on kinetic theory yield Hall viscosities of the same order as the zero field shear viscosity {$\eta_0$, $\etprpt \approx (0.2-0.5) \eta_0$ and  $\etpart \approx (0.2-0.7) \eta_0$} for magnetic fields $B  \approx 10^{19} G$ and $T \approx 300 $ MeV at initialization time. The estimates based on holographic models~\cite{Ammon:2020rvg, Landsteiner:2016stv} predict $\etpart \approx 0.34 \eta_0$ and $\etprpt \approx 0.68 \eta_0$, which is of the same order as the kinetic theory estimates. Our estimates based on kinetic theory and holography both yield similar values for the Hall viscosities, which are in turn comparable to the values of the zero field shear viscosity $\eta_0$. This quantitative agreement between weak- and strong-coupling estimates suggests that Hall viscosity is a robust, largely coupling-insensitive feature of QCD matter under broken time reversal and rotational symmetry.

An important feature of Hall viscosity is that its phenomenological impact is primarily controlled by the magnetic fields at early times in the collision, i.e., on the time scale when the anisotropic flow is generated. 
We quantified how the Hall viscous terms enter the stress tensor at {hydrodynamic initialization time} $\tau_0 \sim 1 fm/c$ using velocity gradients derived from analytic models of pre-equilibrium flow~\cite{Ryu:2021lnx, Alzhrani:2022dpi, Pratt:2008bc, Vredevoogd:2008id, Torrieri:2002jp}. The Hall viscous corrections are comparable in magnitude to the anisotropic viscous stresses and can therefore influence early-time pressure anisotropies, potentially altering the build-up of elliptic~$(v_2)$ and directed flow~$(v_1)$.

Phenomenologically, Hall viscosity induces torque-like couplings between longitudinal and transverse flow harmonics (c.f. Figs.~\ref{figvort} and~\ref{figvortlong}), leading to {rotations of the event-plane phases} $(\Delta \theta, \Delta \phi)$ that could manifest as correlations between global polarization, directed-flow rapidity slopes, and the participant-plane angle~\cite{Rath:2024vyq, PhysRevC.85.044907, starv1s, starv1}. These signatures are experimentally accessible via modern event-engineering and polarization-flow correlators.

Several directions {for future theoretical studies} are: First of all, we would like to quantify the effect of strong vorticity on the Hall transport in the heavy-ion collision~(HIC) which we expect to be of the same order-of-magnitude as the strong magnetic field effect. Another important future direction on the theoretical front is to compute magnetovortical transport coefficients in Veneziano-QCD (V-QCD) at finite temperature and baryon density, in this way testing more realistically the mechanisms for generating Hall viscosity at strong coupling. V-QCD provides a very realistic holographic model of QCD at finite temperature and baryon density.  Since the magnetic field {only
couples to the quark sector}, the naive semiclassical limit in AdS/CFT of large number of colors misses its effect. V-QCD, on the other hand, works in the Veneziano limit, which is a limit of large number of both flavors and colors while maintaining their ratio finite, and thus {accounts for backreaction
from the quark onto the gluon sector}, thereby capturing magnetic-field effects.

Incorporating $\etprpt$ and $\etpart$ into relativistic magneto-hydrodynamic codes will allow a quantitative study of how Hall viscous terms modify freeze-out observables.{\footnote{For a self-consistent (magneto-)hydrodynamic description, also the other anisotropic transport coefficients should be included, most prominently the longitudinal and transverse shear viscosities, as well as the longitudinal and transverse electric conductivities~\cite{Garbiso:2020puw,Cartwright:2021qpp,Cartwright:2022hlg,Amano:2022mlu,Amano:2023bhg,Cartwright:2026bpm,Kaminski:2025ika,Shovkovy:2025yvn,Ghosh:2024fkg,Ghosh:2024owm}. Generically, with increasing anisotropy (stronger magnetic field, stronger rotation) the longitudinal transport coefficient will deviate more from the transverse one.}} The predicted phase-angle shifts  $(\Delta \theta, \Delta \phi)$ could be compared to event-plane correlations such as \eqn{vncorr}, more direct probes of event plane geometry \cite{evcorralice,evcorratlas} and $\Lambda-$hyperon global polarization data. The coupling of shear to spin polarization~\cite{Fu:2021pok, Becattini:2021iol, Becattini:2021suc} suggests a direct relation between Hall viscosity and spin hydrodynamics that warrants investigation. Future global fits could include $\etprpt$ and $\etpart$ as additional transport parameters in Bayesian analyses~\cite{Bernhard:2019bmu,JETSCAPE:2020mzn,JETSCAPE:2020shq,JETSCAPE:2026hdw}. Sensitivity studies using observables such as elliptic flow $v_2$, directed flow $v_1$, polarization correlations, or event-plane tilts would constrain the magnitude and sign of Hall viscosities in the QGP.

\section*{Acknowledgements} We thank {Dmitri Kharzeev for valuable comments and} Aleksi Kurkela for discussions in the early stage of this project, as well as Ante Biladzic and Jiangyong Jia for helpful comments and suggestions on the manuscript. R.M.~acknowledges the support of the German Research Foundation (DFG) through the Collaborative Research Center ToCoTronics, Project-ID 258499086 — SFB 1170, as well as Germany’s Excellence Strategy through the W{\"u}rzburg-Dresden Cluster of Excellence on Complexity and Topology in Quantum Matter - ctd.qmat (EXC 2147, Project-ID 390858490). He furthermore acknowledges hospitality from the Shanghai Institute for Mathematics and Interdisciplinary Sciences (SIMIS) during the final stages of this work, and associated travel support under STCSM Grant 25HB2701900. 
G.T. thanks Bolsa de produtividade CNPQ 305731/2023-8 and FAPESP 2023/06278-2, as well as
participation in the tematico 2023/13749-1 for support.
{M.K.}~was supported, in part, by the U.S.~Department of Energy grant DE-SC0012447. {M.K.}~thanks the Galileo Galilei Institute for Theoretical Physics for the hospitality and the INFN for partial support during the completion of this work. S.M.~thanks the Theoretical Physics III Group at the University of W{\"u}rzburg for their hospitality and support during the early stages of this work.

\appendix

\section{Hall viscosities}\label{App:A}
In 3+1-dimensional fluids described by hydrodynamics the breaking of rotation invariance from $\mathcal{O}(3)$ down to $\mathcal{O}(2)$ is crucial. This can be achieved by introducing a large anisotropy in the plasma, for example a strong magnetic field or global rotation. This leads to different behavior within the plane perpendicular to the anisotropy (e.g. perpendicular to the axis of global rotation) and the planes including the anisotropy direction (e.g. including the axis of rotation). 

In general, the QGP will be subject to at least three anisotropies: magnetic field generated by colliding ions, rotation due to non-zero impact parameter/off-central collision, and anisotropic energy densities as well as pressures due to Bjorken-like expansion of the plasma along the beam line, which we take to be the $z$-direction.  The magnetic field and global rotation axis are aligned with each other, let's assume along the $y$-direction. 
For Hall viscosity it does not matter what type of anisotropy breaks the rotation symmetry, any anisotropy will lead to two distinct Hall viscosities: $\tilde\eta_{||}$ within the plane containing the anisotropy, and $\tilde\eta_\perp$ perpendicular to that plane. 

Also the shear viscosity will split into $\eta_{||}$ and $\eta_\perp$. In holographic systems, only $\eta_\perp$ is required to satisfy $\eta/s=1/(4\pi)$, while $\eta_{||}/s$ generically decreases towards zero when the anisotropy is increased. 
The only known {cases} in which $\eta_{||}/s$ increases with increasing anisotropy are a holographic p-wave superfluid~\cite{Erdmenger:2010xm} {and the charged $\mathcal{N}=4$ Super-Yang-Mills plasma subject to a moderately strong external magnetic field~\cite{Ammon:2020rvg}. Anisotropic shear viscosities along with other anisotropic transport coefficients or anisotropic dispersion relations have been computed holographically in~\cite{Garbiso:2020puw,Cartwright:2021qpp,Cartwright:2022hlg,Amano:2022mlu,Amano:2023bhg,Cartwright:2026bpm,Kaminski:2025ika}.} 

Kubo relations in the case of only one anisotropy, computed for a magnetic field along the $z$-direction, give the transverse Hall viscosity (this is like the one known from condensed matter physics in 2+1-dimensional materials~\cite{Avron_1995,avron1998oddviscosity,Jensen:2011xb,Hoyos:2014pba})
\begin{equation}\label{app:eq:kubo:eta-perp}
    \tilde\eta_\perp = \lim_{\omega\to 0} \lim_{k\to 0} \frac{1}{\omega} \langle T_{xz} (T_{xx}-T_{zz}) \rangle 
\end{equation}
and the longitudinal Hall viscosity (which is not known or discussed in condensed matter literature  
)
\begin{equation}\label{app:eq:kubo:eta-par}
    \tilde\eta_{||} = \lim_{\omega\to 0} \lim_{k\to 0} \frac{1}{\omega}  \langle T_{xy} T_{zy} \rangle 
\end{equation}

In addition to Hall viscosities, there are anisotropic shear viscosities in 3D, denoted by $\eta_\perp$ and $\eta_\parallel$. Their Kubo formulas are

\begin{equation}
    \eta_\perp =  \lim_{\omega\to 0} \lim_{k\to 0} \frac{1}{\omega}  \langle T_{xz} T_{xz} \rangle, 
\end{equation}
and 
\begin{equation}
    \eta_\parallel = \lim_{\omega\to 0} \lim_{k\to 0} \frac{1}{\omega} \langle T_{xy} T_{xy} \rangle   = \lim_{\omega\to 0} \lim_{k\to 0} \frac{1}{\omega}  \langle T_{zy} T_{zy} \rangle,  
\end{equation}
respectively.

\section{The Alekseev mechanism in two-dimensional non-relativistic hydrodynamics}\label{app:alekseev-derivation}

In Ref.~\cite{alekseev2016negative}, it was shown in non-relativistic kinetic theory for 2D electron systems that a Hall viscosity can be induced in a charged plasma by an out-of-plane external magnetic field. The mechanism relies on the Lorentz force exerted by the magnetic field onto the charged particles as they get exchanged between different fluid layers. Consider a electrically charged fluid at zero magnetic field, with kinematic shear viscosity $\nu_0$ given by the relaxation time $\tau_2$ of the second moment of the quasiparticle (in \cite{alekseev2016negative} electron) distribution function via
\begin{equation}
    \nu_0 = \frac{1}{4} v_F^2 \tau_2\,,\quad \frac{1}{\tau_2} = \frac{1}{\tau_{2,ee}} + \frac{1}{\tau_2,0}\,.
\end{equation}
Here $\tau_{2,ee}$ is the relaxation time due to electron-electron interactions, and $\tau_2,0$ due to other effects such as impurities. The latter part will be absent in the quark-gluon plasma.
We include subscript 0 in $\nu_0$ to emphasize that this is shear viscosity in the absence of magnetic field. Ref.~\cite{alekseev2016negative} showed that when switching on a finite but still non-quantizing\footnote{Non-quantizing here means that the magnetic field is weak enough compared to the other scales $\mu$, $T$, such that the Landau level spacing is small compared to the electron-electron interaction rate that is responsible for the hydrodynamic approximation, $\omega_c \tau_{2,ee} \ll 1$ magnetic field $B$, where $\omega_c = \frac{eB}{m}$ is the cyclotron frequency in the non-relativistic regime in SI units\cite{liaokoch}. (In Gaussian units, the Lorentz force differs by a factor of $1/c$, which leads to $\omega_c = \frac{eB}{m c}$.) In the relativistic regime, the mass gets a correction by factor of $\gamma = 1/\sqrt{1 - v^2/c^2}$, thereby, the cyclotron frequency in the relativistic regime in SI units is $\omega_c = \frac{eB}{\gamma m}$.}  magnetic field, both kinematic shear and kinematic Hall viscosities get a correction of form
\begin{equation}\label{eq:Alekseevviscosities}
    \nu_\perp(B) = \nu_{xx}(B) = \frac{\nu_0}{1+(2\omega \tau_2)^2}\,,\quad \tilde{\nu}_\perp (B) = \nu_{xy}(B) = \frac{2 \omega_c\tau_2\nu_0}{1+(2\omega \tau_2)^2}\,.
\end{equation}

The non-relativistic derivation of Ref.~\cite{alekseev2016negative} technically goes as follows.
Let $\vec v(\vec r,t)$ be the velocity field of the fluid flow, and $\tau_2$ the intrinsic relaxation time for the relaxation to equilibrium (related to interactions between the fluid constituents). In kinetic theory, the viscous stress tensor of the fluid per particle and the associated Navier-Stokes equations are given by
\begin{equation}\label{eq:NS}
    m \frac{\partial v_i}{\partial t} = - \frac{\partial \Pi_{ij}}{\partial x_j} - m \frac{v_i}{\tau} + e E_i\,,\quad \Pi_{ij} = m \langle v_i v_j\rangle\,.
\end{equation}
Here $\tau$ is a momentum relaxation timescale, to be included for condensed matter applications but set to zero in the QGP, $E_i$ is an external electric field, and $\langle \rangle$ denotes a statistical average. The Einstein summation convention is assumed. The idea of Ref.~\cite{alekseev2016negative} is now to include, on top of an already thermalized fluid, the effect of the external magnetic field. Since including the magnetic field will change the equilibrium value of $\Pi_{ij}$\footnote{Note $\Pi_{ij}^{0}$ is the local equilibrium value of $\Pi_{ij}$. In other words, $\Pi_{ij}$ relaxes to hydrodynamic form $\Pi_{ij}^{0}$ in timescale $\tau_2$.}, Ref.~\cite{alekseev2016negative} considers the relaxation equation
\begin{equation}\label{eq:relax}
    \frac{\partial \Pi_{ij}}{\partial t} = -\frac{1}{\tau_2}\left(\Pi_{ij} - \Pi_{ij}^{0}\right)\,,\quad 
\end{equation}
In the absence of the magnetic field, the equilibrium solution to \eqref{eq:relax} in the presence of kinematic shear viscosity $\nu_0$ only is 
\begin{equation}
 \Pi^0_{ij} = - m \nu_0 V_{ij}\,,\quad V_{ij} = \frac{\partial V_i}{\partial x_j} + \frac{\partial V_j}{\partial x_i}\,,
\end{equation}
where the average drift velocity $V := \langle v_i \rangle$. We emphasize that $\nu_0$ here is kinematic shear viscosity and not dynamical shear viscosity $\eta_0$ unlike in the main text. The two are related by the relation $\nu \approx  \eta_0/ \varrho$ in the non-relativistic limit, where $\varrho$ is the mass density of the quasiparticles. The Lorentz force due to the magnetic field will now induce additional first and second moments in the velocity distribution,
\begin{eqnarray}\label{eq:moments1}
    \frac{\partial \langle v_i\rangle }{\partial t} &=& \omega_c \epsilon_{zik} \langle v_k \rangle \,,\\\label{eq:moments2}
    \frac{\langle v_i v_j\rangle}{\partial t} &=& \omega_c \left(\epsilon_{zik}\langle v_k v_j\rangle + \epsilon_{zjk}\langle v_i v_k\rangle \right)\,,
\end{eqnarray}
with $\omega_c = \frac{eB}{m}$ the non-relativistic cyclotron frequency. Only \eqref{eq:moments2} will show up on the right hand side of \eqref{eq:relax},
\begin{equation}\label{eq:relax2}
    \frac{\partial \Pi_{ij}}{\partial t} = -\frac{1}{\tau_2}\left(\Pi_{ij} - \Pi_{ij}^{0}\right) + \omega_c \left(\epsilon_{zik}\langle v_k v_j\rangle + \epsilon_{zjk}\langle v_i v_k\rangle \right)\,, 
\end{equation}
leading to a shift in the equilibrium stress tensor
\begin{equation}\label{eq:newequil}
 \Pi_{ij} = \Pi^0_{ij} + \tau_2 \omega_c \left(\epsilon_{zik}\langle v_k v_j\rangle + \epsilon_{zjk}\langle v_i v_k\rangle \right)\,.
\end{equation}
Comparing this to the hydrodynamic form of the {viscous stress tensor} in the presence of Hall viscosity $\nu_H$ and shear viscosity $\nu$, $\Pi_{ij} = m(\nu v_{ij} + \nu_H\epsilon_{zik} v_{kj})$ yields \eqref{eq:Alekseevviscosities}.

The derivation of Ref.~\cite{alekseev2016negative} is strictly only true in the Fermi liquid regime $\mu/T\gg 1$, where $\mu$ is the chemical potential for the electromagnetic $U(1)$ symmetry. For electron fluids, the corrections to the result \eqref{eq:Alekseevviscosities} are not very big, of the order of $20\%$ at most.

\section{Derivation of early-time velocity gradients}\label{app:velocity-grad}
In this appendix, we evaluate various terms appearing in the expressions for velocity gradients of Eq.~\ref{eq:vel-grad}. Recall from Eq.~\eqref{eq:T-Ml-Mn}, we have,
\begin{equation}
    T^{tt} = T^{\tau \tau} + 2 \tau_0 \eta T^{\tau \eta}
\end{equation}
up to linear order in $\eta$. Differentiating and using Eq.~\eqref{eq:Ttautau} and Eq.~\eqref{eq:Ttaueta}, we get,
\begin{align}
    \partial_x T^{tt} & = \left( \cosh y_L(x,y) + 2 \eta \sinh y_L (x,y) \right)\partial_x e(x,y) \nonumber \\
    &+ e(x,y) \left(\sinh y_L(x,y) +2 \eta \cosh y_L (x,y) \right) \partial_x  y_L (x,y)
\end{align}
Substituting this in the universal flow formula, Eq.~\eqref{eq:univ-flow-Pratt}, we get
\begin{align}\label{eq:123}
     \frac{T^{tx}}{T^{tt}} &\approx \frac{\tau_0 \cosh \eta}{2} \frac{\partial_x T^{tt}}{T^{tt}}, \nonumber \\
    & = \frac{\tau_0 \cosh \eta}{2} \Big( \frac{\partial_x e(x,y)}{e(x,y)} + \frac{2 \eta + \tanh y_L (x,y)}{1  +2 \eta \tanh y_L(x,y)} \partial_x y_L(x,y)\Big), \nonumber  \\
    & =  \frac{\tau_0 \cosh \left(\text{arcsinh} \left( \frac{z}{\tau_0} \right) \right)}{2} \Big( \frac{\partial_x e(x,y)}{e(x,y)} + \frac{2 (\text{arcsinh} \left( \frac{z}{\tau_0} \right)) + \tanh y_L (x,y)}{1  +2 (\text{arcsinh} \left( \frac{z}{\tau_0} \right)) \tanh y_L(x,y)} \partial_x y_L(x,y)\Big), 
\end{align}
where we used $\eta = (\text{arcsinh} \left( \frac{z}{\tau_0} \right))$. $\partial_x y_L(x,y)$ and $\partial_x e(x,y)$ appearing in Eq.~\eqref{eq:123} are given by the following expressions
\begin{align}
    \partial_x y_L(x,y) = \frac{f}{2} \tanh(y_{beam}) \Big( \frac{\partial_xT_A(x,y)}{T_A(x,y)} - \frac{\partial_x T_B(x,y)}{T_B(x,y)} \Big)
\end{align}
and
\begin{align}
    \partial_x e(x,y) = \frac{m_N^2 \Big( (T_A(x,y) \cosh(2 y_{beam}) T_B(x,y)) \partial_x T_A(x,y) + (T_B(x,y) + \cosh (2 y_{beam}) T_A(x,y)) \partial_x T_B(x,y) \Big)}{M(x,y) \Big( 2 \text{sinh}(\eta_0) + \sqrt{\pi/2} \sigma_\eta e^{\sigma_\eta^2/2} C_\eta \Big)}
\end{align}

We now derive expressions for the gradients of $\theta$, $y_L$, and $\tanh y_T$ to evaluate velocity gradients of Eq.~\ref{eq:vel-grad}. We have,

\begin{align}
    \partial_x \phi &= -\frac{y}{x^2 + y^2} \\
     \partial_y \phi &= -\frac{x}{x^2 + y^2}
\end{align}
\begin{align}
    \partial_x (\tanh y_T) &= - \frac{\tau_0 \cosh \eta}{2} \Big( \frac{T^{tt} \partial_x^2 T^{tt} - (\partial_x T^{tt})^2}{(T^{tt})^2} \Big) \\
    \partial_y (\tanh y_T) &= - \frac{\tau_0 \cosh \eta}{2} \Big( \frac{T^{tt} \partial_y \partial_x T^{tt} - (\partial_x T^{tt}) (\partial_y T^{tt})}{(T^{tt})^2} \Big) \\
    \partial_z (\tanh y_T) &= - \frac{\tau_0}{2} \cosh \eta \Big ( \frac{T^{tt} \partial_z \partial_x T^{tt} - (\partial_x T^{tt}) (\partial_z T^{tt})}{(T^{tt})^2} \Big) - \frac{\tau_0}{2} \sinh \eta (\partial_z \eta) \frac{\partial_x T^{tt}}{T^{tt}}
\end{align}
Recall $\eta  = \text{arcsinh} \left(z/\tau_0 \right)$, which gives 
\begin{equation}
    \partial_z \eta = \frac{1}{\sqrt{z^2 + \tau_0^2}}
\end{equation}
We obtain following expressions for the gradients of $\tanh y_T$, 

\begin{align}
    \tanh y_T &= \frac{\tau_0}{2} \cosh\eta \Big( \frac{\partial_x e}{e}  + 
     \frac{(\sinh y_L + 2 \eta \cosh y_L ) \partial_x y_L }{(\cosh y_L + 2 \eta \sinh y_L)}
    \Big) 
 \end{align}

\begin{align}
    \partial_x (\tanh y_T) &= -\frac{\tau_0 \cosh \eta}{2} \Big(  \frac{ - (\partial_x e)^2 + e \partial_x^2 e}{e^2} +  \nonumber  \\
    & \frac{(2 - 8 \eta^2) (\partial_x y_L)^2 + (4 \eta \cosh 2 y_L + (1 + 4 \eta^2)\sinh 2 y_L) \partial_x^2 y_L}{2  (\cosh y_L + 2 \eta \sinh y_L)^2}
    \Big) 
\end{align}

\begin{align}
    \partial_y (\tanh y_T) &= -\frac{\tau_0 \cosh \eta}{2} \Big( \frac{- \partial_y e \partial_x e + e \partial_y \partial_x e}{e^2}  +\nonumber \\
    & \frac{2 (1 - 4 \eta^2) \partial_x y_L \partial_y y_L + (4 \eta \cosh 2 y_L + (1 +4 \eta^2)\sinh 2 y_L) \partial_y \partial_x y_L}{2  (\cosh y_L + 2 \eta \sinh y_L)^2}
    \Big) 
\end{align}

\begin{align}
    \partial_z (\tanh y_T) &= \frac{\tau_0}{2} \partial_z \eta \Big( - \frac{\sinh \eta \partial_x e}{e} -  \nonumber \\
    & \frac{(2 \cosh \eta + 2 \eta  \cosh 2 y_L \sinh \eta  + (1 + 4 \eta^2) \cosh y_L \sinh y_L \sinh \eta ) \partial_x y_L}{(\cosh y_L + 2 \eta \sinh y_L)^2}
    \Big) 
\end{align}

Whereas, using the expression for $y_L$ in Eq.~\eqref{eq:yL}, the gradients of $y_L$ come out to be,

\begin{align}
    \partial_x y_L &= f \; \frac{\text{tanh}( y_{\text{beam}})}{2} \Big( \frac{\partial_x T_A}{T_A} - \frac{\partial_x T_B}{T_B} \Big)
\end{align}

\begin{align}
    \partial_y y_L &= f \; \frac{\text{tanh}( y_{\text{beam}})}{2} \Big( \frac{\partial_y T_A}{T_A} - \frac{\partial_y T_B}{T_B} \Big)
\end{align}

\begin{align}
    \partial_x^2 y_L &= f \; \frac{\text{tanh}( y_{\text{beam}})}{2}  \Big( \frac{- (\partial_x T_A)^2 + T_A \partial_x^2 T_A}{T_A^2}  + \frac{ (\partial_x T_B)^2 - T_B \partial_x^2 T_B}{T_B^2}\Big)
\end{align}

\begin{align}
    \partial_y \partial_x y_L &=  f \; \frac{\text{tanh}( y_{\text{beam}})}{2} \Big( \frac{- ( \partial_x T_A) ( \partial_y T_A) + T_A (\partial_y \partial_x T_A)}{T_A^2} + \frac{ ( \partial_x T_B) ( \partial_y T_B) - T_B (\partial_y \partial_x T_B)}{T_B^2} \Big)
\end{align}

Finally we evaluate gradients of $e$. Recall from Eq.~\eqref{eq:e=Ne},
\begin{align}
    e(x,y) &= \mathcal{N}_e(x,y) = \frac{M(x,y)}{\kappa}, \nonumber \\
    \kappa &= 2 \text{sinh}(\eta_0) + \sqrt{\pi/2} \sigma_\eta e^{\sigma_\eta^2/2} C_\eta, \nonumber \\
    C_\eta &= e^{\eta_0} \text{erfc}\left( -\sqrt{\frac{1}{2}} \sigma_\eta \right) + e^{-\eta_0} \text{erfc} \left( \sqrt{\frac{1}{2}} \sigma_\eta \right)
\end{align}
The expression for $M(x,y)$ is given in Eq.~\eqref{eq:Mxy}. We get the following expressions for the gradients,
\begin{align}
    \partial_x e(x,y) &= \frac{\partial_x M(x,y)}{\kappa } \\
    \partial_y e(x,y) &= \frac{\partial_y M(x,y)}{\kappa }  \\
    \partial^2_x e(x,y) &= \frac{\partial^2_x M(x,y)}{\kappa } \\
    \partial_y \partial_x e(x,y) &= \frac{\partial_y \partial_x M(x,y)}{\kappa } 
\end{align}

\begingroup
\sloppy
\bibliography{literatur}
\endgroup

\end{document}